\pdfoutput=1

\documentclass[twocolumn,floatfix,showpacs,superscriptaddress]{revtex4}
\usepackage{graphicx}
\usepackage{amssymb,amsmath}
\begin{document}

\title{
  Conformal mapping and shot noise in graphene
}

\author{Adam Rycerz}
\affiliation{Marian Smoluchowski Institute of Physics, 
Jagiellonian University, Reymonta 4, PL--30059 Krak$\acute{o}$w, Poland}
\affiliation{Institut f\"{u}r Theoretische Physik, 
Universit\"{a}t Regensburg, D--93040, Germany}
\author{Patrik Recher}
\affiliation{Institut f\"{u}r Theoretische Physik und Astrophysik, 
Universit\"{a}t W\"{u}rzburg, \\Am Hubland, D--97074 W\"{u}rzburg, Germany}
\author{Michael Wimmer}
\affiliation{Institut f\"{u}r Theoretische Physik, 
Universit\"{a}t Regensburg, D--93040, Germany}

\begin{abstract}
Ballistic transport through a collection of quantum billiards in undoped
graphene is studied analytically within the conformal mapping technique. 
The billiards show \emph{pseudodiffusive} behavior, with the conductance 
equal to that of a classical conductor characterized by the conductivity 
$\sigma_0=4e^2/\pi h$, and the Fano factor $F=1/3$. By shrinking at least one
of the billiard openings, we observe a \emph{tunneling} behavior,
where the conductance shows a power-law decay with the system size, and the 
shot noise is Poissonian ($F=1$). In the crossover 
region between tunneling and pseudodiffusive regimes, the conductance 
$G\approx (1-F)\times se^2/h$. The degeneracy $s=8$ for the Corbino disk,
which preserves the full symmetry of the Dirac equation, $s=4$ for billiards 
bounded with smooth edges which break the symplectic symmetry, and $s=2$
when abrupt edges lead to strong intervalley scattering. An alternative, 
analytical or numerical technique, is utilized for each of the billiards
to confirm the applicability of the conformal mapping for various boundary 
conditions.
\end{abstract}

\date{\today}
\maketitle

\section{Introduction}

The isolation of single layers of carbon (graphene) whose low-energy spectrum
is described by the Dirac-Weyl Hamiltonian of massless spin-$1/2$ fermions 
\cite{Gei07}, has offered physicists the unique possibility to test the 
predictions of relativistic quantum mechanics in a condensed phase.
A particular attention focuses on ballistic transport \cite{Imr96}, as the
unusual band structure of a carbon monolayer \cite{Wal47} leads simultaneously 
to a divergent Fermi wavelength $\lambda_F\!\rightarrow\!\infty$ in the undoped 
graphene limit, and to a zero bandgap. 
For these reasons, the quantum-mechanical wave character of an electron 
plays an essential role in transport even through a macroscopic graphene 
sample, provided that the influence of disorder is negligible \cite{Nom07}.
A separate issue concerns the fact that Dirac fermions in graphene occur in 
two degenerate families, resulting from the presence of two different 
valleys in the band structure. 
This valley degree of freedom offers conceptually new possibilities to control
charge carriers---the so-called ``valleytronics'' \cite{Ryc07}.

So far, extensive theoretical studies of ballistic transport, based on
mode-matching analysis for the Dirac equation \cite{Kat06a,Two06}, are 
available 
for a \emph{rectangular} graphene sample of width $\mathcal{W}$, length 
$\mathcal{L}$, and various types of boundary conditions. In the regime of large
aspect ratios $\mathcal{W}/\mathcal{L}\gg 1$, the conductance of an undoped 
sample scales as $G=\sigma_0\mathcal{W}/\mathcal{L}$, 
with the universal conductivity $\sigma_0=4e^2/\pi h$, regardless of boundary 
conditions \cite{Ryc08}. Moreover, as shown by Tworzyd{\l}o \emph{et al.} 
\cite{Two06}, the Fano factor in this case coincides with that of a diffusive
wire ($F=1/3$). Also, the transmission eigenvalues of these two systems 
display the same  distribution. This analogy coined the term of 
\emph{pseudodiffusive transport}, which describes ballistic graphene 
properties in the universal conductivity limit.

Recent experiments report an agreement with the theoretical predictions of 
Refs.\ \cite{Kat06a,Two06} for either the conductance \cite{Mia07} or the Fano 
factor \cite{Dan08}. Furthermore, the temperature  dependence of the 
conductivity \cite{Du08} also shows an approximate agreement with the ballistic 
theory generalized to finite temperatures \cite{Mue08}. 
However, even for low temperatures, the convergence 
with $\mathcal{W}/\mathcal{L}\rightarrow\infty$ is much slower than predicted. 
In particular, for the largest aspect ratio $\mathcal{W}/\mathcal{L}=24$ 
studied in Ref.\ \cite{Dan08} the deviations from the limiting values 
$G\mathcal{L}/\mathcal{W}=\sigma_0$ and $F=1/3$ are close to 
$10\%$, whereas results of Ref.\ \cite{Two06} show the convergence should be 
already reached for moderate aspect ratios $\mathcal{W}/\mathcal{L}\gtrsim 4$. 
A clear explanation of this discrepancy is missing, but it is usually 
attributed to the fact that boundary conditions used in theoretical works, 
which describe either an abrupt termination of a perfect honeycomb lattice or 
an infinite mass confinement \cite{Akh08}, may not model the real-sample edges 
correctly \cite{Gol07}.

In this work, we consider ballistic graphene systems of geometries for which 
the boundary effects are absent or suppressed. The paper is organized as 
follows: In Sec.\ \ref{conmap} we briefly recall the mode-matching analysis for
a graphene strip, and show how to employ the conformal symmetry of the Dirac 
equation for undoped graphene \cite{Kat08} to obtain analytically the 
transmission eigenvalues for other systems. Then, in Sec.\ \ref{corbdi} the 
method is applied to the Corbino disk. The results are compared with those
obtained by direct mode-matching for angular momentum eigenstates, a relation
with the nonrelativistic electron gas in the disk setup is also discussed. 
In Sec.\ \ref{bimass} we study two basic billiards bounded with mass 
confinement: a~finite section of the Corbino disk and a quantum dot with 
circular edges. The results obtained with the conformal mapping technique are 
confirmed by the computer simulation of transport using the tight-binding model
on a honeycomb lattice. 
Finally, in Sec.\ \ref{ribb} we study numerically the transport 
\emph{across} an infinitely long nanoribbon by utilizing the $4$-terminal 
recursive Green's function algorithm \cite{Wim08a}, as well as across a finite 
section of a~nanoribbon with an abrupt lattice termination. All the systems 
show pseudodiffusive transport behavior in a wide range of geometrical 
parameters. 
A further analogy between them appears when (at least) one of the leads is 
narrow in comparison to the characteristic length of the conducting region $L$.
Namely, the conductance in such a~\emph{quantum tunneling} regime shows 
a~power-law decay $G\propto L^{-\alpha}$, where $\alpha$ is a nonuniversal 
(geometry-dependent) exponent. Moreover, it is related to the shot noise by 
$F\approx 1-Gh/(se^2)$, so that the Poissonian limit ($F=1$) is approached for 
large $L$. The symmetry-dependent factor $s=8$ in the presence of full spin, 
valley and symplectic degeneracy (the case of the Corbino disk), or $s=4$ when 
the mass confinement breaks symplectic symmetry of the Dirac equation. A final 
reduction to $s=2$ may be reached by adding abrupt (i.e.\ armchair) edges, 
which scatter the valleys.

The original feature of the geometries studied in this article is that the 
influence of sample edges are eliminated (for the Corbino setup) or irrelevant,
as the spatial current distribution is not uniform, but concentrated far away 
from the edges.  This is why we believe that our theoretical findings could be 
confirmed experimentally with better precision than that for rectangular 
samples, as the role of boundary conditions is strongly suppressed.

\section{Transport of Dirac fermions and conformal mapping}
\label{conmap}

\begin{figure}[!ht]
\centerline{\includegraphics[width=\linewidth]{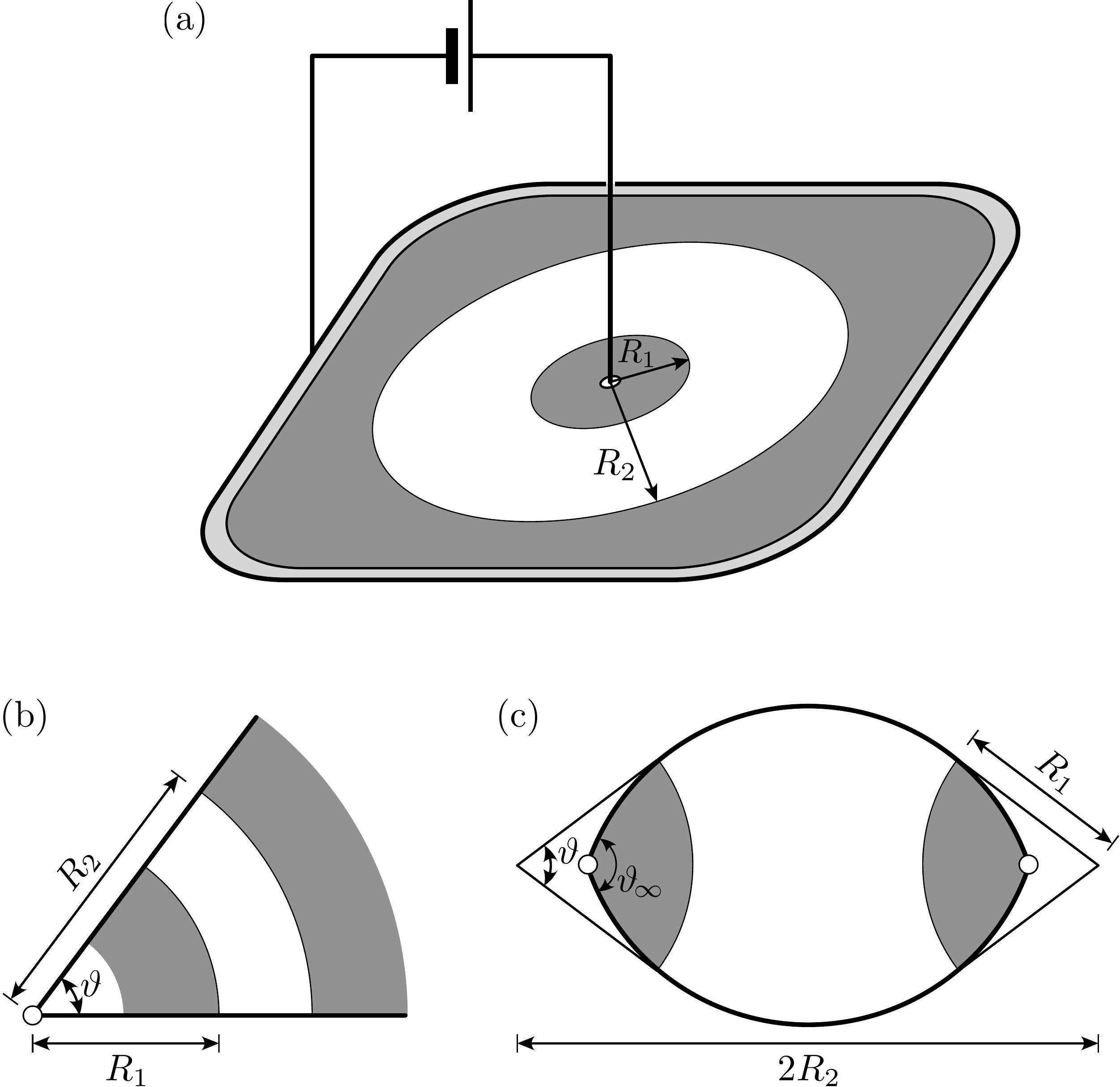}}
\caption{\label{corbcido}
Quantum billiards in undoped graphene studied analytically (schematic). 
(a) The Corbino disk with inner radius $R_1$, and outer radius $R_2$.
(b) Generic section of the disk characterized by the spanning angle 
$\vartheta$. (c) Quantum dot with circular edges. A voltage source,
shown on panel (a) only, drives the current through each of the devices.
Shadow areas on all panels mark heavily-doped (so highly-conducting) 
graphene leads, white dots are the poles of conformal transformation mapping 
a given system onto a strip of Fig.\ \ref{interf12}.
Thick lines on panels (b,c) indicate infinite-mass confinement.
}
\end{figure}

The compact derivation of transmission eigenvalues of a weakly-doped (or 
undoped) graphene sample coupled to heavily-doped graphene leads is known due 
to Sonin \cite{Son08}, who pointed out that one can first calculate the 
reflection and transmission amplitudes for an interface between weakly-doped 
and heavily-undoped regions, and then employ the double-contact formula 
\cite{Dat97}. Here we show that the derivation of Ref.\ \cite{Son08} can
be easily adopted to the Corbino disk, a finite section of it, and to a quantum 
dot with circular edges (all shown in Fig.\ \ref{corbcido}), as these systems 
can be obtained from a strip by appropriate conformal transformations.

\begin{figure}[!ht]
\centerline{\includegraphics[width=0.65\linewidth]{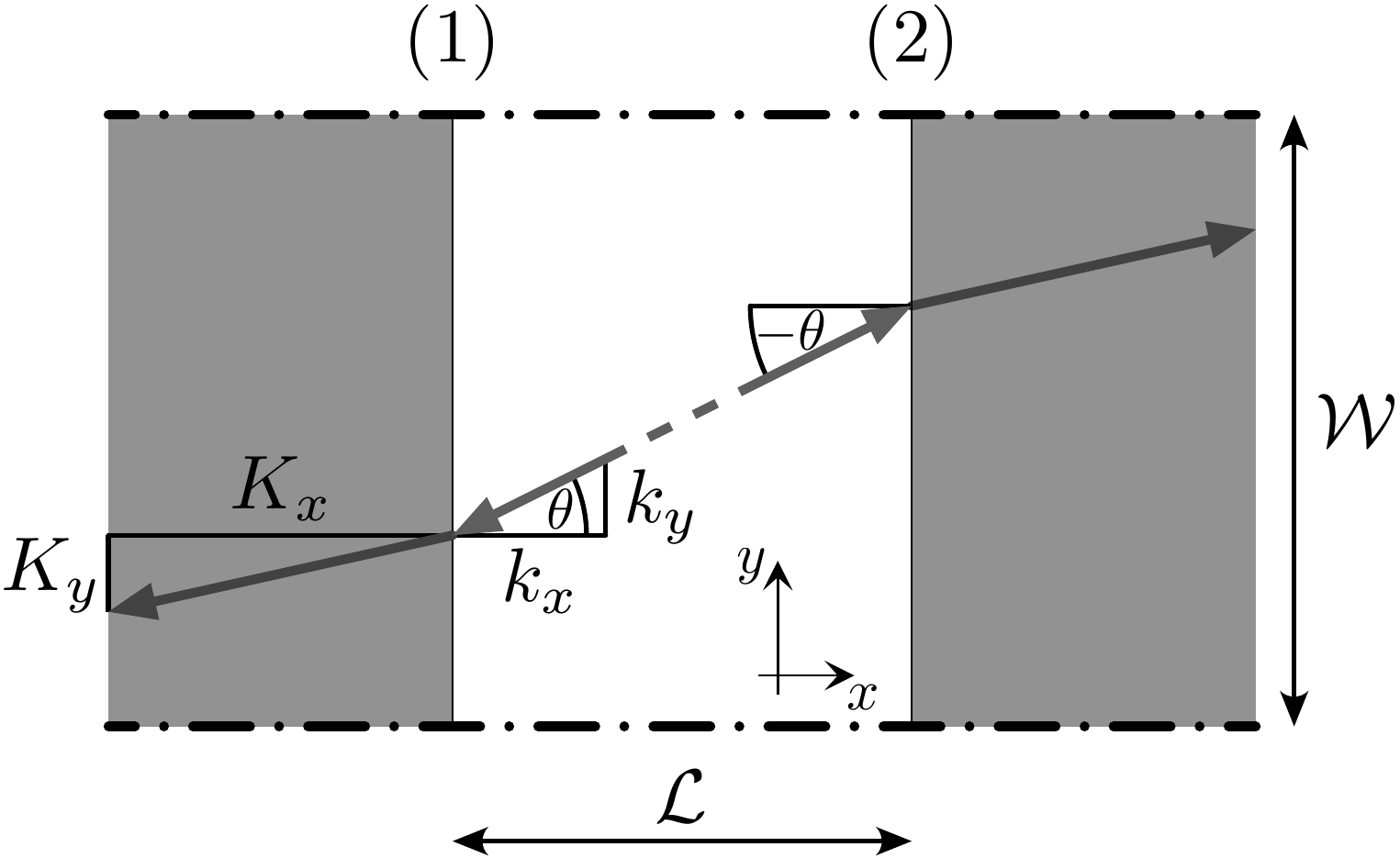}}
\caption{\label{interf12}
Scattering on interfaces (1) and (2) between weakly-doped (white area) and 
heavily-doped (shaded area) regions in graphene. Horizontal dashed-dot lines
mark symbolically generic boundary conditions applied to a strip.
}
\end{figure}

\subsection{Mode-matching for a graphene strip}
Let us first consider an electron crossing from the weakly-doped region 
($x>0$) to the heavily-doped one ($x<0$), as depicted in Fig.\ \ref{interf12}.
The Dirac Hamiltonian for graphene has the well-known form \cite{Ben08}
\begin{equation}
\label{dirham}
  H_0=v_F\mbox{\boldmath$\sigma$}\cdot\mbox{\boldmath$p$},
\end{equation}
where $v_F$ is the Fermi velocity, 
$\mbox{\boldmath$\sigma$}=(\sigma_x,\sigma_y)$ 
is the vector operator build of Pauli matrices for the sublattice-pseudospin
degree of freedom, and $\mbox{\boldmath$p$}=-i\hbar(\partial_x,\partial_y)$ is 
the in-plane momentum operator. Due to translational invariance along the 
$y$-axis, the solution of the Dirac equation with energy $E=\hbar v_{F}k$ may 
be written as $\Psi(x,y)=\chi_\theta(x)e^{ik_yy}$ \cite{TraInv}, with the 
transverse momentum $k_y=K_y$ ($k_i$ and $K_i$ with $i=x,y$ denote momentum 
components in the weakly- and heavily-doped regions, respectively), 
and the spinor
\begin{equation}
\chi_\theta(x)=\left\{\begin{array}{cc}
  \left(\!\begin{array}{c}1\\ -e^{-i\theta}\end{array}\!\right)e^{-ik_xx}
   + r_1\left(\begin{array}{c}1\\ e^{i\theta}\end{array}\right)e^{ik_xx},
    & x\!>\!0 \\
  t_1\sqrt{\frac{k_x}{k}}\left(\begin{array}{c}1\\ -1\end{array}\right)
  e^{-iK_xx},  & x\!<\!0
\end{array}\right.
\end{equation}
where $e^{i\theta}=(k_x+ik_y)/k$, and the limit of an infinite doping ($k\ll K$) 
is imposed. The continuity of the two spinor components on both sides of the 
interface leads to expressions for the reflection and transmission amplitudes
\begin{equation}
\label{rt1th}
  r_1=\frac{e^{-i\theta}-1}{e^{i\theta}+1},
  \ \ \ \ \ \ 
  t_1=\frac{2\sqrt{\cos\theta}}{e^{i\theta}+1}.
\end{equation}
The amplitudes $r_1$ and $t_1$ depend solely on the angle of incidence 
$\theta$ (see Fig.\ \ref{interf12}), what illustrates the generic feature
of ballistic transport in graphene that is insensitive to the lead details
\cite{Sch07}. The reflection and 
transmission amplitudes for an electron crossing from the undoped region to 
the second heavily-doped lead are $r_2=r_1^\star$ and $t_2=t_1^\star$ (up to a 
phase factor), as the angle of incidence $\theta\rightarrow -\theta$ in this 
case. Thus, employing the double-contact formula of Ref.\ \cite{Dat97} the
total transmission probability for phase-coherent transport through the system 
of Fig.\ \ref{interf12} is
\begin{equation}
\label{tthphi}
  T=
  \frac{\left|t_1t_2\right|^2}{\left|1-r_1r_2e^{2i\phi_{12}}\right|^2}=
  \frac{1}{1+(\tan\theta\sin\phi_{12})^2},
\end{equation}
where $\phi_{12}\equiv\int_1^2k_xdx$ is the phase-shift earned by an electron 
when passing from the first interface to the second one. 
The above expression holds true for either propagating modes (for which 
$k_x=\sqrt{k^2-k_y^2}$) or, as an analytic continuation, for evanescent modes 
($k_x=iq_x$, with $q_x=\sqrt{k_y^2-k^2}$). For a confined geometry, the 
quantization of transverse momenta is determined by boundary conditions 
\cite{Kat06a,Two06}. Namely, $k_y=k_y^l$ (with $l$-integer) may be written
in a compact form 
\begin{equation}
\label{kyquant}
  k_y^l=\frac{ g\pi(l+\gamma) }{ \mathcal{W} }, 
\end{equation}
where $g=1,2$ for the closed and (generalized) periodic b.c., 
respectively. $\gamma=\frac{1}{2}$ for either mass confinement or 
antiperiodic b.c.\ studied in this paper. (For a nanotube-like geometry as
considered in Ref.\ \cite{Kat06a}, $\gamma=0$ corresponds to periodic b.c.)

\subsection{Transmission via evanescent modes}
Here we limit ourselves to the case of undoped graphene ($k\rightarrow 0$), 
in which the charge transport is carried fully by evanescent modes. 
An analytic continuation yields $\tan\theta\rightarrow i$ and 
$
  \phi_{12} \rightarrow ig\pi j\mathcal{L}/\mathcal{W}
$
in Eq.\ (\ref{tthphi}), where we use the quantization (\ref{kyquant}) and 
define the half-integer $j\equiv l+\frac{1}{2}$.
As pointed out by Katsnelson and Guinea \cite{Kat08}, the zero-energy
solution of the Dirac equation may be obtained via conformal transformation
that links the considered geometry to a simple one, for which the wavefunction 
is known \cite{Dircon}. In particular, if the conformal transformation
$z(w)$ turns the system under consideration into a rectangle of width 
$\mathcal{W}$ and length $\mathcal{L}$ (Fig.\ \ref{interf12}), the 
transmission probability for the $j$-th evanescent mode may be written as
\begin{equation}
\label{tjlambda}
  T_j=\frac{1}{\cosh^2\left[gj\ln\Lambda\{z(w)\}\right]}
  =\frac{4}{\left(\Lambda^{gj} + \Lambda^{-gj}\right)^2},
\end{equation}
where $j=\pm\frac{1}{2},\pm\frac{3}{2},\dots$ (with the degeneracy 
$T_j=T_{-j}$). 
Notice that the amplitudes (\ref{rt1th}) remain unchanged after 
applying an arbitrary conformal transformation to the coordinate system of 
Fig.\ \ref{interf12}, so the only term in Eq.\ (\ref{tthphi}) affected by 
the transformation $z(w)$ is the phase-shift 
$\phi_{12}\rightarrow ig j\ln\Lambda$.
The real functional $\Lambda\{z(w)\}$ is defined by
\begin{equation}
\label{lambdadef}
  \ln\Lambda\{z(w)\}\equiv\pi\mathcal{L}/\mathcal{W}.
\end{equation}
The explicit form of $\Lambda\{z(w)\}$ depends on the geometry, and is 
given below for the examples of conformal transformation $z(w)$ having 
one and two poles in a complex plane, which allows us to obtain expressions
for transmission probabilities through a finite section of the Corbino disk
and through a quantum dot with circular edges, respectively.

But first, we discuss the two basic physical regimes of quantum transport in 
graphene, which are described by opposite limits of Eq.\ (\ref{tjlambda}).
The conductance of undoped graphene \cite{Kat06a,Two06} is given by the 
Landauer formula
\begin{equation}
\label{gsumtj}
  G=\frac{se^2}{h}\sum_{j=\frac{1}{2},\frac{3}{2},\dots} T_j=
  s\pi\sigma_0\sum_{j}\left(\Lambda^{gj}+\Lambda^{-gj}\right)^{-2},
\end{equation}
with the degeneracy $s=4$ (spin and valley) for smooth mass confinement, and
$s=8$ for antiperiodic b.c.\ due to an additional (symplectic) symmetry 
\cite{perifoo}. The universal conductivity is $\sigma_0\equiv 4e^2/\pi h$.
The Fano factor also follows from summing over the modes
\begin{equation}
\label{fsumtj}
  F=\frac{\displaystyle\sum_{j=\frac{1}{2},\frac{3}{2},\dots}T_j(1-T_j)}%
{\displaystyle\sum_{j=\frac{1}{2},\frac{3}{2},\dots}T_j},
\end{equation}
but is affected by the symmetry-dependent factors $(g,s)$ only via $T_j$-s 
(\ref{tjlambda}).

For the limit $\ln\Lambda\ll 1$, we can replace summation in Eq.\ 
(\ref{gsumtj}) by integration, and get
\begin{equation}
\label{gdifflam}
  G\approx G_\mathrm{diff}=\frac{\pi\sigma_0}{\ln\Lambda\{z(w)\}},
\end{equation}
where we use the relation $s=4g$, valid for the two classes of b.c.\ 
studied here. In the $\ln\Lambda\ll 1$ limit, the relevant information about 
transmission probabilities is given by their statistical distribution
\begin{equation}
  \rho(T)= \frac{2}{T\sqrt{1-T}}\frac{G_\mathrm{diff}}{\pi\sigma_0}.
\end{equation}
As the distribution $\rho(T)$ coincides with the known distribution 
\cite{Dor84} for diffusion modes in a disordered metal, $\ln\Lambda\ll 1$
constitutes a \emph{pseudodiffusive} regime of transport through graphene 
billiards. Notice that the generic conformal transformation $z(w)$ affects 
$\rho(T)$ only via the prefactor $G_\mathrm{diff}$. In particular, the Fano 
factor
\begin{equation}
\label{favertt}
  F=1-\frac{\langle T^2\rangle}{\langle T\rangle}\approx \frac{1}{3},
\end{equation}
regardless of the particular form of $z(w)$. This observation may also help to
understand why experimental results \cite{Dan08} generally show better 
agreement with theory for the Fano factor rather than for the conductance.
For instance, various geometrical defects (such as a corrugation of the 
lead-graphene interface) may affect $G_\mathrm{diff}$ strongly, but not affect 
$F$ at all.

In the opposite limit ($\ln\Lambda\gg 1$), we find from (\ref{tjlambda}) that
$T_{1/2}\gg T_{3/2}\gg\dots$, leading to
\begin{equation}
\label{gftunnlam}
  G\approx s\pi\sigma_0\Lambda^{-g}, \ \ \ \ \ \ 
  F\approx 1-G\frac{h}{se^2}.
\end{equation}
These expressions constitute
a \emph{quantum-tunneling} regime for ballistic graphene, in which the 
transport is governed by a single electronic mode with the four-fold
(spin and valley) degeneracy. Below, we provide examples illustrating how
the power-law dependence of $G$ on $\Lambda$ may be followed by a power-law
decay of $G$ with the characteristic length-scale of the system.

\section{Application to the Corbino disk}
\label{corbdi}

The Corbino setup, in which the graphene sample formed as an annulus is 
attached to coaxial leads, as shown schematically in Fig.\ \ref{corbcido}(a), 
seems to be the simplest way to eliminate boundary effects, which are claimed
to strongly affect experimental results for rectangular samples with small and
moderate aspect ratios \cite{Mia07,Dan08}. In this section, we first utilize
the conformal mapping technique to find transmission eigenvalues for an 
\emph{undoped} disk, and then compare the results with that obtained by a 
direct wave-function matching, possible also for a \emph{doped} disk.

\begin{figure}[!ht]
\centerline{\includegraphics[width=\linewidth]{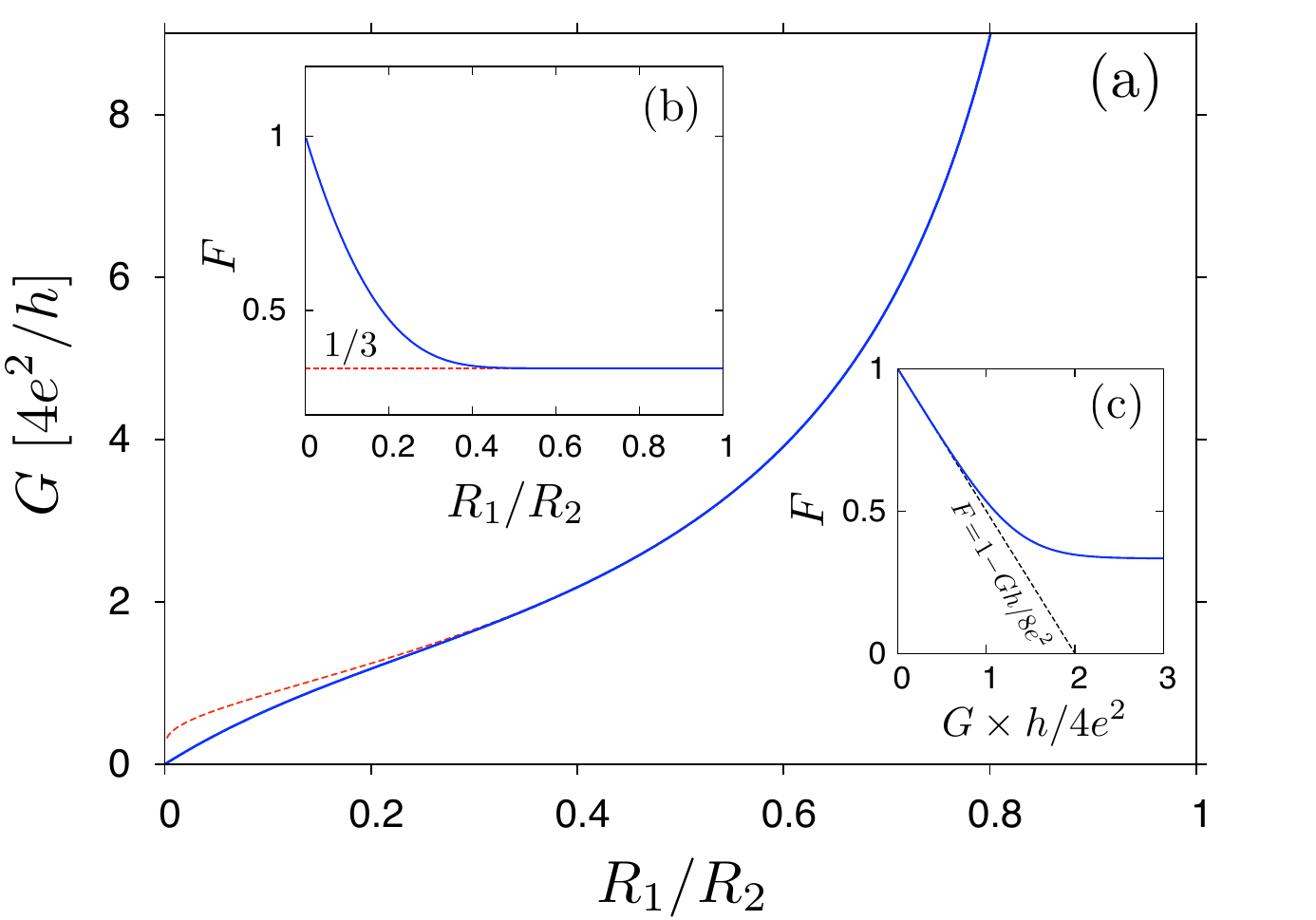}}
\caption{\label{gfdiscr}
Conductance and Fano factor for the undoped Corbino disk in graphene, as 
a function of the radii ratio (a,b) and the shot noise vs conductance diagram 
(c). The curves calculated from Eqs.\ (\ref{gsumtj}) and (\ref{fsumtj}) are 
plotted with solid lines on all panels. Dashed lines show the pseudodiffusive 
limit (\ref{gfdiffdsk})  on panels (a,b) and the tunneling limit 
(\ref{gftunndsk}) on panel (c). 
}
\end{figure}

\subsection{Conformal mapping for an undoped disk}
The conformal transformation that changes the Corbino disk 
with the inner radius $R_1$ and the outer radius $R_2$, shown in Fig.\ 
\ref{corbcido}(a), into a rectangle of the width $\mathcal{W}$ and the length 
$\mathcal{L}$ (see Fig.\ \ref{interf12}), is given by \cite{Smy68}
\begin{equation}
\label{contradsk}
  z = \frac{\mathcal{W}}{2\pi}\mbox{Log}\frac{w}{R_1}.
\end{equation}
(Hereinafter, we use the symbol $\mbox{Log}$ to denote 
the natural logarithm in a complex domain.)
For the complex variable $z=x+iy$, with $0\leqslant x\leqslant\mathcal{L}$,
and $0\leqslant y\leqslant\mathcal{W}$, transformation (\ref{contradsk}) leads
to $R_1\leqslant|w|\leqslant R_2$ and $0\leqslant\arg w\leqslant 2\pi$ 
provided the condition $R_2/R_1=e^{2\pi\mathcal{L}/\mathcal{W}}$ is satisfied. 
Using (\ref{lambdadef}), such a condition implies the functional 
$\Lambda\{z(w)\}$ to have the form  
\begin{equation}
\label{lacontradsk}
   \Lambda=\Lambda(R_1,R_2)=\left(\frac{R_2}{R_1}\right)^{1/2}.
\end{equation}

As the conformal mapping is known, the only part to be explained now
are the boundary conditions applied to a strip of Fig.\ \ref{interf12}.
To define them, one needs to notice that after a rotation by $2\pi$ in the 
coordinate system of Fig.\ \ref{corbcido}(a), the spinor part of the 
wavefunction acquires the Berry phase \cite{Ull08,Cas09} $e^{i\pi\sigma_z}=-1$. 
Within the mapping (\ref{contradsk}), a rotation by $2\pi$ turns into a shift 
along the $y$-axis in Fig.\ \ref{interf12} by a strip width $\mathcal{W}$.
This is why the spinor-rotational invariance of the original wavefunction 
implies antiperiodic boundary conditions 
$\Psi(x,y+\mathcal{W})=-\Psi(x,y)$ for a strip. Such boundary conditions, 
together with the functional $\Lambda\{z(w)\}$ given by (\ref{lacontradsk}), 
lead the formula (\ref{tjlambda}) for transmission probabilities to a~form 
\begin{equation}
\label{tjlambdadsk}
  T_j=\frac{1}{\cosh^2\left[j\ln\left(R_1/R_2\right)\right]},\ \ \ \ \ \ 
  j={\textstyle\frac{1}{2},\frac{3}{2},\frac{5}{2},\dots}.
\end{equation}
A generalization for the setup with circular, but not coaxial contacts
is presented in Appendix~\ref{appcir}.

The dependence of the conductance (\ref{gsumtj}) and the Fano factor 
(\ref{fsumtj})
on the radii ratio $R_1/R_2$ is plotted in Fig.\ \ref{gfdiscr} (solid lines).
The limiting behavior for $R_1/R_2\approx 1$, corresponding $\ln\Lambda\ll 1$
(\ref{lacontradsk}) is characterized by $G\approx G_\mathrm{diff}$ 
(\ref{gdifflam}), with
\begin{equation}
  \label{gfdiffdsk}
  G_\mathrm{diff}=\frac{2\pi\sigma_0}{\ln\left(R_2/R_1\right)}, \ \ \ \ \ \ 
  \mbox{and}\ \ \ \ \ \ F\approx\frac{1}{3}.
\end{equation}
The formula for $G_\mathrm{diff}$ coincides with the well-known classical 
conductance of the Corbino disk \cite{Shi97}. 
The asymptotic values (\ref{gfdiffdsk}) are depicted with dashed red lines 
on Fig.\ \ref{gfdiscr}(a,b). In the opposite limit ($R_1\ll R_2$), 
Eq.\ (\ref{gftunnlam}) takes the form
\begin{equation}
  \label{gftunndsk}
  G\approx 8\pi\sigma_0\frac{R_1}{R_2}, \ \ \ \ \ \ 
  F\approx 1-G\frac{h}{8e^2}.
\end{equation}
The second formula from above is shown in Fig.\ \ref{gfdiscr}(c) with dashed 
black line.

The results presented in Fig.\ \ref{gfdiscr} show that the pseudodiffusive 
formulas (\ref{gfdiffdsk}) for $G$ and $F$ match the exact expressions 
(\ref{gsumtj}) and (\ref{fsumtj}) with $T_j$ given by (\ref{tjlambdadsk}) 
in a relatively wide range of ratios $R_1/R_2$. 
Namely, the agreement becomes better than $1\%$ if $R_1\geqslant 0.29R_2$ 
for the conductance, and if $R_1\geqslant 0.43R_2$ for the Fano factor. 
For smaller $R_1/R_2$, one can
identify the crossover from the pseudodiffusive to quantum-tunneling behavior.
In particular, the exact values of $G$ are closer to the tunneling formula 
(\ref{gftunndsk}) than to the pseudodiffusive formula (\ref{gfdiffdsk}) 
below $R_1/R_2= 0.11$. The same is observed for $F$ below $R_1/R_2=0.16$. 
The most characteristic feature of the tunneling regime is the relation 
$G\approx (1-F)\times 8e^2/h$, following from (\ref{gftunndsk}). It
is satisfied with an accuracy better than $10\%$ for $G\lesssim 4e^2/h$ (or 
$F\gtrsim 0.5$), corresponding to $R_1/R_2\lesssim 0.2$. In this range, we
also find that the conductance decays (at fixed $R_1$) as $G\propto 1/L$, where 
$L\equiv R_2-R_1\approx R_2$ is the characteristic length of the sample area. 

\begin{figure}[!ht]
\centerline{\includegraphics[width=\linewidth]{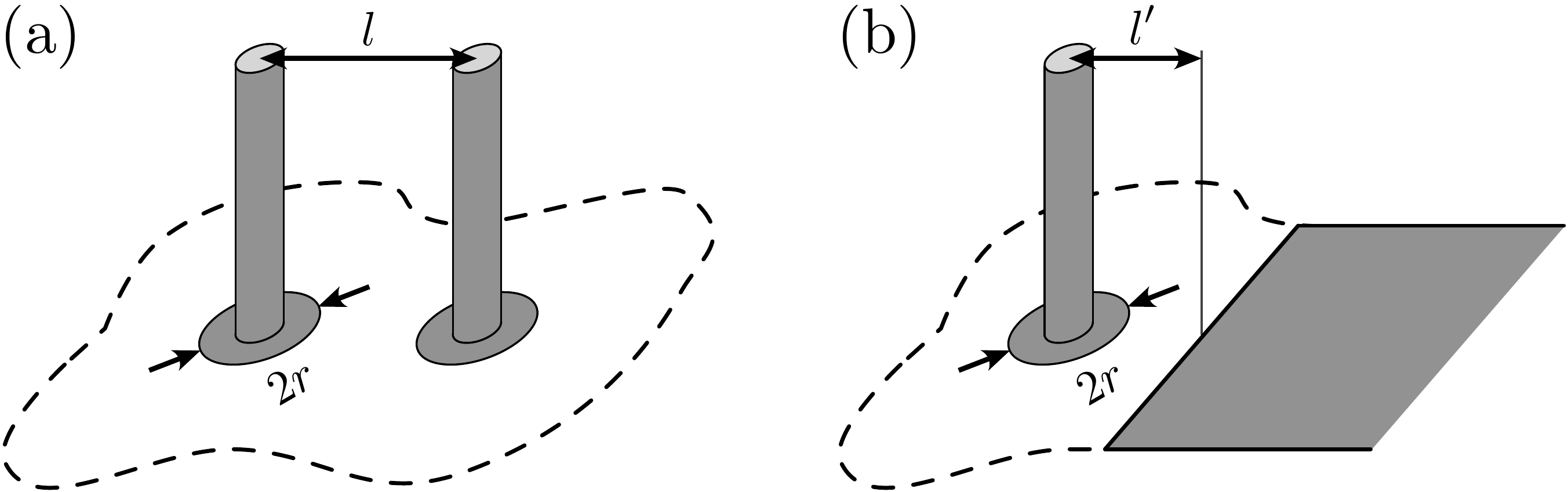}}
\caption{\label{flak2prob}
  Large graphene flake probed by two circular leads of radius $r$ separated 
  by the distance $l$ (a), and by a lead placed in the distance $l'$ from a 
  straight interface between undoped and heavily-doped regions (b).
}
\end{figure}

A similar, power-law decay of the conductance with the sample length is 
predicted for geometries with non-coaxial leads, considered in 
Appendix~\ref{appcir}. In the two limiting situations, the M\"{o}bius 
transformation (\ref{mobius}) maps an infinite plane (hemiplane) with two (one)
narrow circular openings onto the Corbino disk. Physically, these two 
situations correspond to the setup consisting of two circular leads probing 
a~large graphene sample (see Fig.\ \ref{flak2prob}a), and of one circular lead 
and a long straight interface between the undoped and the heavily-doped region
playing the role of a second lead (see Fig.\ \ref{flak2prob}b).
In the first case, the mapping (\ref{mobius}) leads to $\Lambda\approx l/r$ 
and, subsequently, to the quadratic decay of the conductance
\begin{equation}
\label{gtunnflak2a}
  G\approx 8\pi\sigma_0\left(\frac{r}{l}\right)^2\ \ \ \ \mbox{for}\ \  r\ll l
\end{equation}
(with the radius of each lead $r$ and the distance between leads $l$). 
In the second case, the functional $\Lambda\approx (2l'/r)^{1/2}$ and the 
conductance 
\begin{equation}
\label{gtunnflak2b}
  G\approx 4\pi\sigma_0\frac{r}{l'} 
\end{equation}
shows reciprocal decay with the sample area length, similarly as observed for 
the Corbino disk. The approximate relation between the conductance and the 
Fano factor $G\approx (1-F)\times 8e^2/h$ holds true for both situations of
Fig.\ \ref{flak2prob}, showing the tunneling-transport regime in graphene
appears generically for a setup consisting of (at least) one narrow circular 
lead.

\subsection{Electron transport at finite doping}

\begin{figure}[!ht]
\centerline{\includegraphics[width=0.8\linewidth]{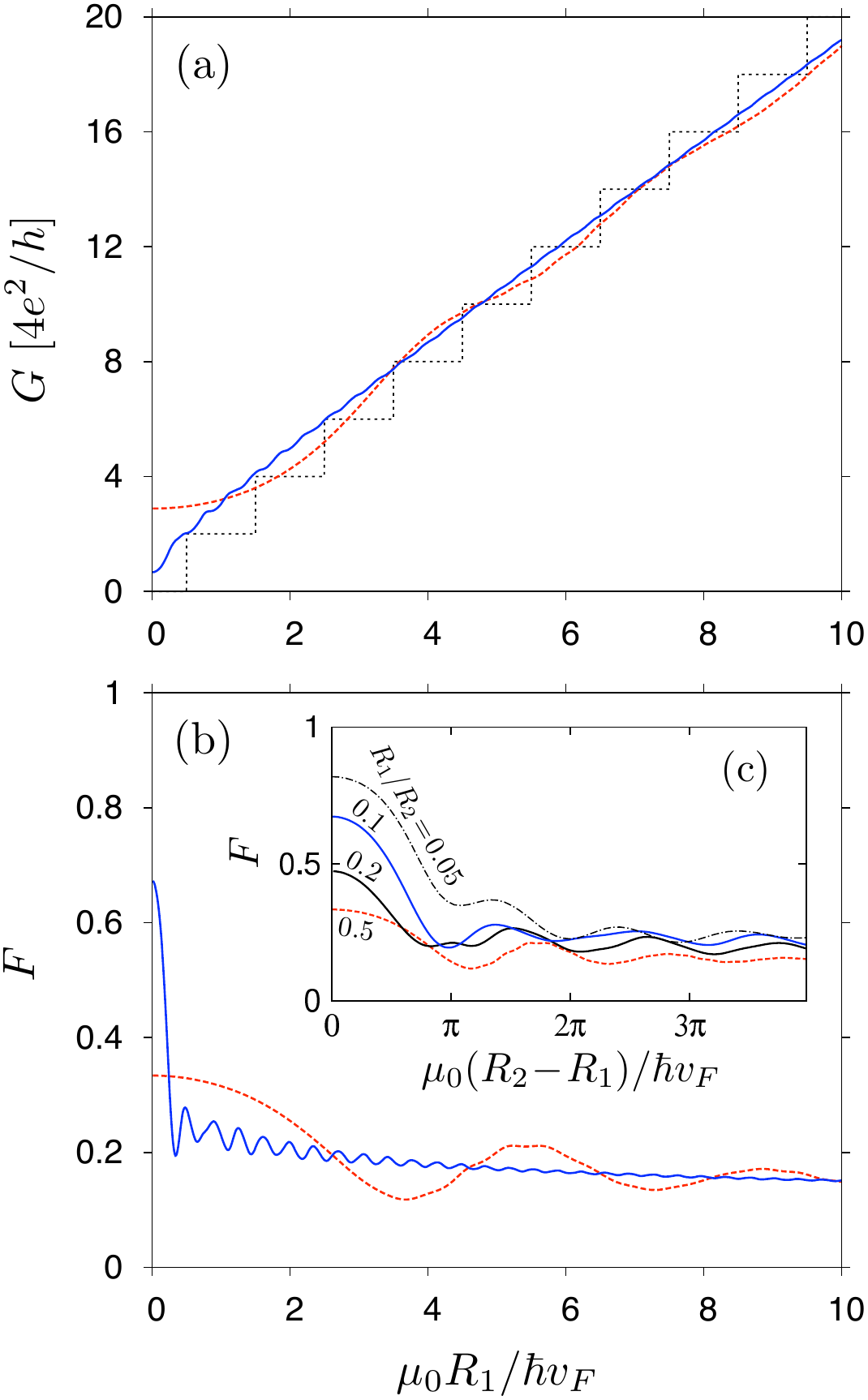}}
\caption{\label{gfdisckr12}
  Chemical potential dependence of the conductance (a) and Fano factor (b,c) 
  at a fixed radii ratio $R_1/R_2$. Solid and dashed lines on panels (a,b) 
  correspond to $R_1/R_2=0.1$ and $0.5$, respectively. 
  The dotted line on panel (a) is the 
  semi-classical approximation for the conductance. Panel (c) shows the Fano
  factor as a function of the chemical potential in the units of $\hbar v_F/
  (R_2\!-\!R_1)$, with $R_1/R_2$ specified for each curve on the plot. 
}
\end{figure}

We complement the study of the Corbino disk in graphene with its transport 
properties at finite doping, characterized by the chemical potential $\mu_0
\equiv\pm\hbar v_Fk$ (where $\mu_0>0$ and $\mu_0<0$ refers to electron and hole
doping, respectively). The analysis is closely related to that for the 
electronic levels of graphene rings \cite{Rec07}. The single valley Hamiltonian
for the doped disk reads
\begin{equation}
\label{dirhamcir}
  H=H_0+U(r)\sigma_0,
\end{equation}
where $H_0$ is given by (\ref{dirham}), the electrostatic potential $U(r)=U_0$ 
if $R_1\leqslant r\leqslant R_2$, and $U(r)=U_\infty$ otherwise. 
The chemical potential $\mu_0=E-U_0$ in the 
disk, or $\mu_\infty=E-U_\infty$ in the leads. 
The rotational invariance of the problem allows us to perform the 
mode-matching for each eigenstate of the total angular momentum $J_z=l_z+
\hbar\sigma_{z}/2$ (with $l_z\equiv -i\hbar\partial_\varphi$ the orbital angular momentum) 
separately. The eigenstate of the Hamiltonian (\ref{dirhamcir}) corresponding
to the $j$-th eigenvalue of $J_z$ can be written as
\begin{multline}
\label{jeigen}
  \psi_j(r,\varphi)=e^{i(j-1/2)\varphi}\left(\begin{array}{c}
    \chi_{j,A}(r) \\ \chi_{j,B}(r)e^{i\varphi}
  \end{array}\right)\\=e^{i(j-\sigma_z/2)\varphi}\chi_j(r),
\end{multline}
where $j$ is a half-odd integer $j=\pm\frac{1}{2},\pm\frac{3}{2},\dots$.
For the electron doping ($E-U(r)>0$), the radial components $\chi\equiv 
[\chi_A,\chi_B]^T$ for the incoming and outgoing waves are given 
(up to the normalization) by
\begin{equation}
\label{chiinoutr}
  \chi_j^{\mathrm{in}}=\left(\begin{array}{c}
    H_{j-1/2}^{(2)}(\rho) \\ iH_{j+1/2}^{(2)}(\rho)
  \end{array}\right),\ \ \ 
  \chi_j^{\mathrm{out}}=\left(\begin{array}{c}
    H_{j-1/2}^{(1)}(\rho) \\ iH_{j+1/2}^{(1)}(\rho)
  \end{array}\right),
\end{equation}
where  $H_{\nu}^{(1,2)}(\rho)$ is the Hankel function of the (first,second) kind, 
and the dimensionless radial coordinate is $\rho=|E-U(r)|r/(\hbar v_F)$ (so 
$\rho=kr$ in the disk and $\rho=Kr$ in the leads, with $K\equiv|\mu_\infty|/
(\hbar v_F)$). The radial current density is $j_{r}=ev_F\left<
\psi_j^{\mathrm{in}(\mathrm{out})}|\sigma_r|\psi_j^{\mathrm{in}(\mathrm{out})}\right>=\pm 
4ev_F/(\pi\rho)$, where the upper (lower) sign is for $\psi_j^{\mathrm{in}}$ 
($\psi_j^{\mathrm{out}}$), $\sigma_r=\sigma_x\cos\varphi+\sigma_y\sin\varphi$, and 
we use the identity $\mbox{Im}[H_{\nu}^{(1)}(\rho)H_{\nu+1}^{(2)}(\rho)]=
2/(\pi\rho)$. 
For the hole doping ($E-U(r)<0$), the wavefunctions are 
$\tilde{\chi}_j^{\mathrm{in}(\mathrm{out})}=\left[\chi_j^{\mathrm{in}(\mathrm{out})}
\right]^{\star}$, where we use the relation $H_{\nu}^{(2)}=\left[H_{\nu}^{(1)}
\right]^{\star}$. The transmission and reflection amplitudes are obtained by 
wavefunction matching at $r=R_1$ and $r=R_2$. (Note that the $\varphi$ 
dependence of the spinor (\ref{jeigen}) plays no role for the mode-matching 
analysis.)

Details of the calculations are given in Appendix~\ref{appmod}. For 
$|\mu_\infty|\rightarrow\infty$ (the heavily-doped leads limit), the 
transmission probability for the $j$-th mode $T_j=T_j(\mu_0)$ reads
\begin{equation}\label{tjmodma}
T_j=\frac{16}{\pi^2k^2R_1R_2}\,\frac{1}{(\mathfrak{D}_j^+)^2+(\mathfrak{D}_j^-)^2},
\end{equation}
with
\begin{multline}\label{djmodma}
  \mathfrak{D}_j^\pm=\mbox{Im}\left[H_{j-1/2}^{(1)}(kR_1)H_{j\mp 1/2}^{(2)}(kR_2)\right. \\ \left.\pm H_{j+1/2}^{(1)}(kR_1)H_{j\pm 1/2}^{(2)}(kR_2)\right].
\end{multline}
Eqs.\ (\ref{gsumtj},\ref{fsumtj}) for $G$ and $F$ remain unchanged, since we 
again observe the symmetry $T_{-j}=T_j$. In addition, the particle-hole symmetry
$T_j(-\mu_0)=T_j(\mu_0)$ allows us to limit the discussion to $\mu_0>0$.

Numerical values for the conductance and Fano factor of the doped
disk are presented in Fig.\ \ref{gfdisckr12}. Following the idea of Kirczenow 
\cite{Kir94}, we compare (in Fig.\ \ref{gfdisckr12}a) the exact quantum 
conductance given by Eqs.\ (\ref{gsumtj}) and (\ref{tjmodma}) 
with the semiclassical approximation for large angular momenta
\begin{equation}
\label{gsemiclass}
  G_{\mbox{\scriptsize s-cl}}=\frac{8e^2}{h}\left(j_1+\frac{1}{2}\right),
\end{equation}
where $j_1=\mbox{int}(kR_1-\frac{1}{2})+\frac{1}{2}$ is the maximal value 
of $j$ such that $j\leqslant kR_1$. Surprisingly, the quantization steps of 
$G_{\mbox{\scriptsize s-cl}}$ (dotted black line) are missing in the actual data even
for an extremely small radii ratio (solid blue and dashed red line for 
$R_1/R_2=0.1$ and $0.5$, respectively). Instead, weak modulation with a period
$\approx\!\pi\hbar v_F/(R_2-R_1)$ is observed when varying $\mu_0$.
Earlier, conductance quantization (with the steps of $4e^2/h$) was 
predicted to appear for a graphene strip with a moderate aspect ratio 
$\mathcal{W}/\mathcal{L}\lesssim 1$ \cite{Ryc07,Ryc08}. 
The quantization with the steps of $8e^2/h$
was found theoretically for a bipolar junction in graphene, which shows the 
Goos-H\"{a}nchen effect \cite{Bee09}. The lack of conductance quantization
observed here for the Corbino disk shows the role of evanescent modes showing 
a slow (power-law) decay with distance is crucial also far away from the Dirac 
point, what exhibits a~striking consequence of the angular-momentum 
conservation.

Similar to the strip geometry \cite{Two06}, the conductance minimum at 
$\mu_0=0$ corresponds to the maximum of the Fano factor (see Figs.\ 
\ref{gfdisckr12}b). The peak width shrinks approximately as 
$\pi\hbar v_F/(R_2-R_1)$ (for more datasets, plotted as a function of 
$\mu_0(R_2-R_1)/\hbar v_F=k(R_2-R_1)$, see Fig.\ \ref{gfdisckr12}c). 
From an analytical treatment of the limit $\mu_{0}\rightarrow 0$
for angular-momentum eigenstates (see Appendix~\ref{appmod}), we find that 
Eq.\ (\ref{tjlambdadsk}) for $T_j$, obtained within the conformal mapping 
technique, is reproduced.

\subsection{Comparison with the Schr\"{o}dinger system}

\begin{figure}[!ht]
\centerline{\includegraphics[width=0.95\linewidth]{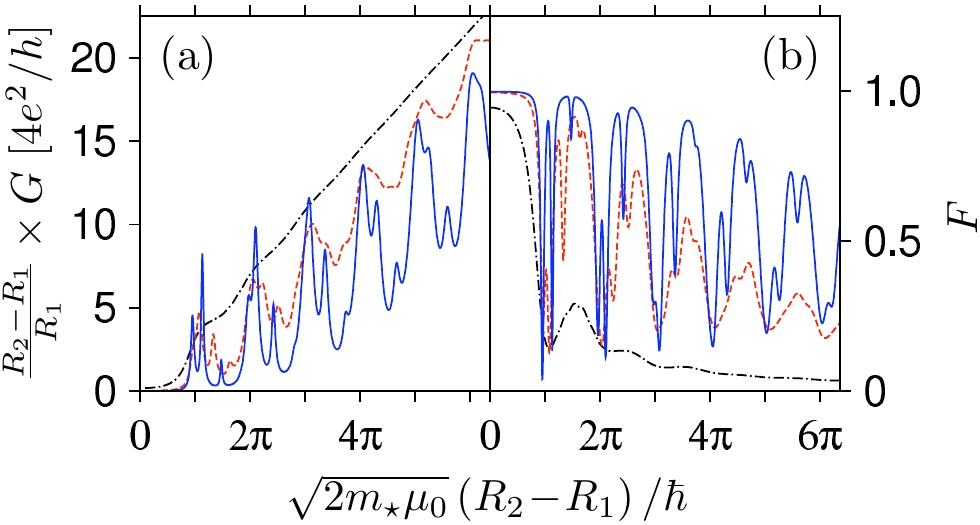}}
\caption{\label{gf2deg}
  Chemical potential dependence of the conductance (a) and the Fano factor (b) 
  for the Corbino disk in a~2DEG. Different lines in each panel correspond to 
  different values of the radii ratio: $R_1/R_2=0.1$ (solid blue lines), $0.2$ 
  (dashed red lines), and $0.1$ (black dash-dotted lines). The electrostatic
  potential step is fixed at $\sqrt{2m_{\star}(U_0-U_\infty)}R_1/\hbar=7$.
}
\end{figure}

As the Corbino disk containing Dirac fermions, described by the Hamiltonian 
(\ref{dirhamcir}), has not been studied in the literature yet, a comparison 
with the corresponding Schr\"{o}dinger system is desirable for the sake of 
completeness. 
The existing theoretical works \cite{Kir94}, however, focus on the model with
a special, angular-momentum dependent effective potential which simplifies the 
analysis, but makes a~relation to the Dirac system studied here unclear. For
this reason, we now present a mode-matching analysis for
two-dimensional nonrelativistic electron gas (2DEG) arranged in a Corbino setup
with the same potential profile as applied to Dirac fermions earlier in this 
paper. 

The Schr\"{o}dinger equation for the Corbino disk in a~2DEG reads
\begin{equation}
\left[-\frac{\hbar^2}{2m_\star}\nabla^2+U(r)\right]\Psi = E\Psi,
\end{equation}
where $m_\star$ is the effective mass, and the electrostatic potential $U(r)$ 
is chosen identically as in the Hamiltonian (\ref{dirhamcir}). The solutions
are written in the form of orbital-momentum $l_z$ eigenstates
\begin{equation}
  \Psi_l(r,\varphi)=e^{il\varphi}\Phi_l(r),
\end{equation}
with $l$ integer, and the radial wavefunction $\Phi_l(r)$ a complex scalar.
The propagating modes in the leads exist only for $\mu_\infty>0$, and have the 
form $\Phi_l^\mathrm{in}(r)=H_l^{(2)}(Kr)$ and $\Phi_l^\mathrm{out}(r)=H_l^{(1)}(Kr)$,
where $K\equiv\sqrt{2m_{\star}\mu_\infty/\hbar^2}$, and we assume scattering from 
the outer lead. For the disk area, two linearly independent solutions are 
given by $H_l^{(1)}(kr)$ and $H_l^{(2)}(kr)$ (with $k\equiv\sqrt{2m_{\star}|\mu_0|/
\hbar^2}$) for $\mu_0>0$. Otherwise, the solutions are given by modified Bessel 
functions $I_l(kr)$ and $K_l(kr)$.
The mode-matching analysis is carried out separately for each value of $l$ 
\cite{match2deg}, leading to the transmission propabilities
\begin{equation}
\label{ttl2deg}
  T_l=\frac{1}{|\mathfrak{M}_l|^2}\left(\frac{64}{\pi^2K^2R_1R_2}\right)^2,
\end{equation}
where
\begin{multline}
  \mathfrak{M}_l=\mathcal{F}_l^{(1,1)}(K,k,R_2)\mathcal{F}_l^{(2,2)}(K,k,R_1) \\
  -\mathcal{F}_l^{(1,2)}(K,k,R_2)\mathcal{F}_l^{(2,1)}(K,k,R_1),
\end{multline}
\begin{multline}
  \mathcal{F}_l^{(i,j)}(K,k,r)=\left[H_{l-1}^{(i)}(Kr)-H_{l+1}^{(i)}(Kr)\right]\mathcal{C}_l^{(j)}(kr)\\-(k/K)H_{l}^{(i)}(Kr)\left[\mathcal{C}_{l-1}^{(j)}(kr)\mp\mathcal{C}_{l+1}^{(j)}(kr)\right],
\end{multline}
with $i,j=1,2$, and the upper (lower) sign corresponding to $\mu_0>0$ 
($\mu_0<0$). We further define
\begin{eqnarray}\label{calcdef}
  && \mathcal{C}_l^{(1)}=\Theta(\mu_0)H_l^{(1)}+\Theta(-\mu_0)I_l, \nonumber \\
  && \mathcal{C}_l^{(2)}=\Theta(\mu_0)H_l^{(2)} + 
  \frac{4}{\pi}(-)^l\Theta(-\mu_0)K_l,
\end{eqnarray}
with the step function $\Theta(x)=1$ for $x>0$ or $\Theta(x)=0$ otherwise.

Numerical values of the conductance and the Fano factor following from Eq.\ 
(\ref{ttl2deg}) are presented in Fig.\ \ref{gf2deg} \cite{land2deg} for a~large
but finite value of the doping in the leads, adjusted such that $\sqrt{2m_{\star}
(U_0-U_\infty)}R_1/\hbar=7$. Both $G$ and $F$ are plotted as functions of 
$k(R_2-R_1)=\sqrt{2m_{\star}\mu_0}(R_2-R_1)/\hbar$ for fixed values of the radii 
ratio $R_1/R_2=0.1$, $0.2$ and $0.5$ (solid, dashed, and dash-dot lines, 
respectively); $G$ is additionally rescaled by a factor $(R_2-R_1)/R_1$ to 
illustrate its asymptotic behavior for $kR_1\gg 1$, which is insensitive to the 
ratio $R_1/R_2$ \cite{KirGscl}. We also limit the discussion to $\mu_0>0$,
as the propabilities $T_l$ given by Eq.\ (\ref{ttl2deg}) decay rapidly for 
$\mu_0<0$, due to lack of propagating modes in the sample area.

Although the values of $G$ shown in Fig.\ \ref{gf2deg}(a) are close to the
semiclassical result \cite{KirGscl}, the quantization steps are absent in the
data. Instead, we observe Fabry-P\'{e}rot oscillations with the amplitude
increasing with $R_2/R_1$, for either $G$ or $F$. We attribute the conductance
quantization, reported in earlier works \cite{Kir94} to the particular choice
of the effective radial potential (note that the existing experiments for 
the Corbino disk in a~2DEG \cite{Tay98} found no conductance quantization).
The main difference in transport through the Corbino disk geometry between 
massless fermions in graphene and massive fermions in a~2DEG, is the reduced 
backscattering at the contacts and the absence of details of the leads (i.e.\ 
the doping) in the former case \cite{infdop2deg}. This is a direct consequence 
of the energy-independent velocity in graphene, which is also responsible for 
the Klein-tunneling phenomena \cite{Kat06b}. Moreover, we note a~suppression 
of the Fabry-P\'{e}rot oscillations for the relativistic system.

\section{Quantum billiards bounded with smooth edges}
\label{bimass}

\subsection{Section of the disk and circular quantum dot}
A simple generalization of the formula (\ref{contradsk}) leads to the 
conformal transformation that changes a finite section of the Corbino disk 
with the inner radius $R_1$, the outer radius $R_2$, and the spanning angle 
$\vartheta$ (shown in Fig.\ \ref{corbcido}b) into a rectangle of the width 
$\mathcal{W}$ and the length $\mathcal{L}$, which is given by
\begin{equation}
\label{contra1}
  z = \frac{\mathcal{W}}{\vartheta}\mbox{Log}\frac{w}{R_1}.
\end{equation}
For $z=x+iy$, where $0\leqslant x\leqslant\mathcal{L}$ and 
$0\leqslant y\leqslant\mathcal{W}$, we get $R_1\leqslant|w|\leqslant R_2$ and 
$0\leqslant\arg w\leqslant\vartheta$ (with $0<\vartheta<2\pi$), under the 
condition that $R_2/R_1=e^{\vartheta\mathcal{L}/\mathcal{W}}$. Using (\ref{lambdadef}), 
such a condition leads to the functional $\Lambda\{z(w)\}$ in the form  
\begin{equation}
\label{lacontra1}
   \Lambda=\Lambda(R_1,R_2,\vartheta)=\left(\frac{R_2}{R_1}\right)^{\pi/\vartheta}.
\end{equation}
Thus, substituting (\ref{lacontra1}) into (\ref{gsumtj}) leads to the exact 
expression for the conductance of a section of the Corbino disk. Notice that 
the transmission probabilities $T_j$ for the full disk (\ref{tjlambdadsk}) 
are \emph{not} reproduced for $\vartheta=2\pi$, as the mass confinement is now 
present in the system. Instead, they are equal for $\vartheta=\pi$, what causes 
the conductance of such a half-disk to be equal to half of the full disk 
conductance for arbitrary $R_1/R_2$. The pseudodiffusive limit $\ln\Lambda\ll 
1$ is realized for $R_1\approx R_2$, and the conductance (\ref{gdifflam}) is
\begin{equation}
\label{gcordif}
  G\approx G_\mathrm{diff}=
  \frac{\sigma_0\vartheta}{\ln\left(R_2/R_1\right)}.
\end{equation}
The above formula coincides with (\ref{gfdiffdsk}) for $\vartheta=2\pi$.
The opposite, quantum tunneling limit ($\ln\Lambda\gg 1$) is reached for 
$R_1\ll R_2$  where formula (\ref{gftunnlam}) gives
\begin{equation}
  G\approx 4\pi\sigma_0\left(\frac{R_1}{R_2}\right)^{\pi/\vartheta}.
\end{equation}
In this case, the conductance decays (at fixed $R_1$) with the 
characteristic length $L\approx R_2$ as $G\propto L^{-\pi/\vartheta}$. The 
reciprocal decay, observed in Section \ref{corbdi} for the full disk, now
appears at $\vartheta=\pi$.

As a next example, we consider the conformal transformation, which changes
the quantum dot shown in Fig.\ \ref{corbcido}(c)
into a rectangle of the width $\mathcal{W}$ and the length $\mathcal{L}$. 
The transformation is given by the formula \cite{Smy68}
\begin{equation}
\label{contra2}
  z-z_0=\frac{\mathcal{W}}{\vartheta_\infty}\mbox{Log}\frac{w+r}{w-r},
\end{equation}
with the condition $(R_2-R_1+r)^2/(R_2-R_1-r)^2=
e^{\vartheta_\infty\mathcal{L}/\mathcal{W}}$, which leads to
\begin{equation}
\label{lacontra2}
\Lambda(R_1,R_2,\vartheta)=
\left(\frac{r-R_1+R_2}{r+R_1-R_2}\right)^{2\pi/\vartheta_\infty}.
\end{equation}
The origin of the coordinate system of Fig.\ \ref{interf12} is now
shifted to $z_0\equiv(\mathcal{L}+i\mathcal{W})/2$. The poles
of the transformation (marked by white dots in Fig.\ \ref{corbcido}b)
are placed at $w=\pm r$, with $r\equiv\sqrt{R_2^2-R_1^2}$. The angle 
$\vartheta_\infty=\vartheta_\infty(R_1,R_2,\vartheta)$, at which the dot edges 
intersect each other, is
\begin{equation}
\label{thinf}
  \vartheta_\infty=2\pi\Theta(\vartheta-\vartheta_0)-\mbox{sgn}
  (\vartheta-\vartheta_0)\xi(\vartheta,\vartheta_0),
\end{equation}
\begin{multline}
  \mbox{with}\ \ \ \ \ 
  \xi(\vartheta,\vartheta_0)=
  2\arcsin\left(\frac{\sin\frac{\vartheta}{2}\sin\frac{\vartheta_0}{2}}%
{1-\cos\frac{\vartheta}{2}\cos\frac{\vartheta_0}{2}}\right), \hfill
\end{multline}
and $\vartheta_0\equiv 2\arccos\left(R_1/R_2\right)$. Again, substituting
(\ref{lacontra2}) into (\ref{gsumtj}) provides one with the exact expression
for the system conductance, which reaches the pseudodiffusive limit for 
$R_1\approx R_2$, where $G\approx G_\mathrm{diff}$ [see Eq.\ (\ref{gdifflam})]
and
\begin{equation}
\label{gcirdif}
  G_\mathrm{diff}=\frac{\sigma_0\vartheta_\infty}{\ln
  \left[(R_2-R_1+r)^2/(R_2-R_1-r)^2\right]},
\end{equation}
whereas for the quantum tunneling limit $R_1\ll R_2$, the formula 
(\ref{gftunnlam}) reads
\begin{equation}
\label{gcirtun}
  G\approx 4\pi\sigma_0\left(
    \frac{r+R_1-R_2}{r-R_1+R_2}\right)^{2\pi/\vartheta_\infty}.
\end{equation}
This leads to an asymptotic form 
$G\sim (R_1/R_2)^{2\pi/\vartheta}$, as $\vartheta_\infty\rightarrow\vartheta$ and
$r\rightarrow R_2$ for $R_1/R_2\rightarrow 0$, while the Fano factor 
approaches the Poissonian value $F\approx 1$.

\subsection{Numerical results}
\label{numres}

\begin{figure}[!ht]
\centerline{\includegraphics[width=\linewidth]{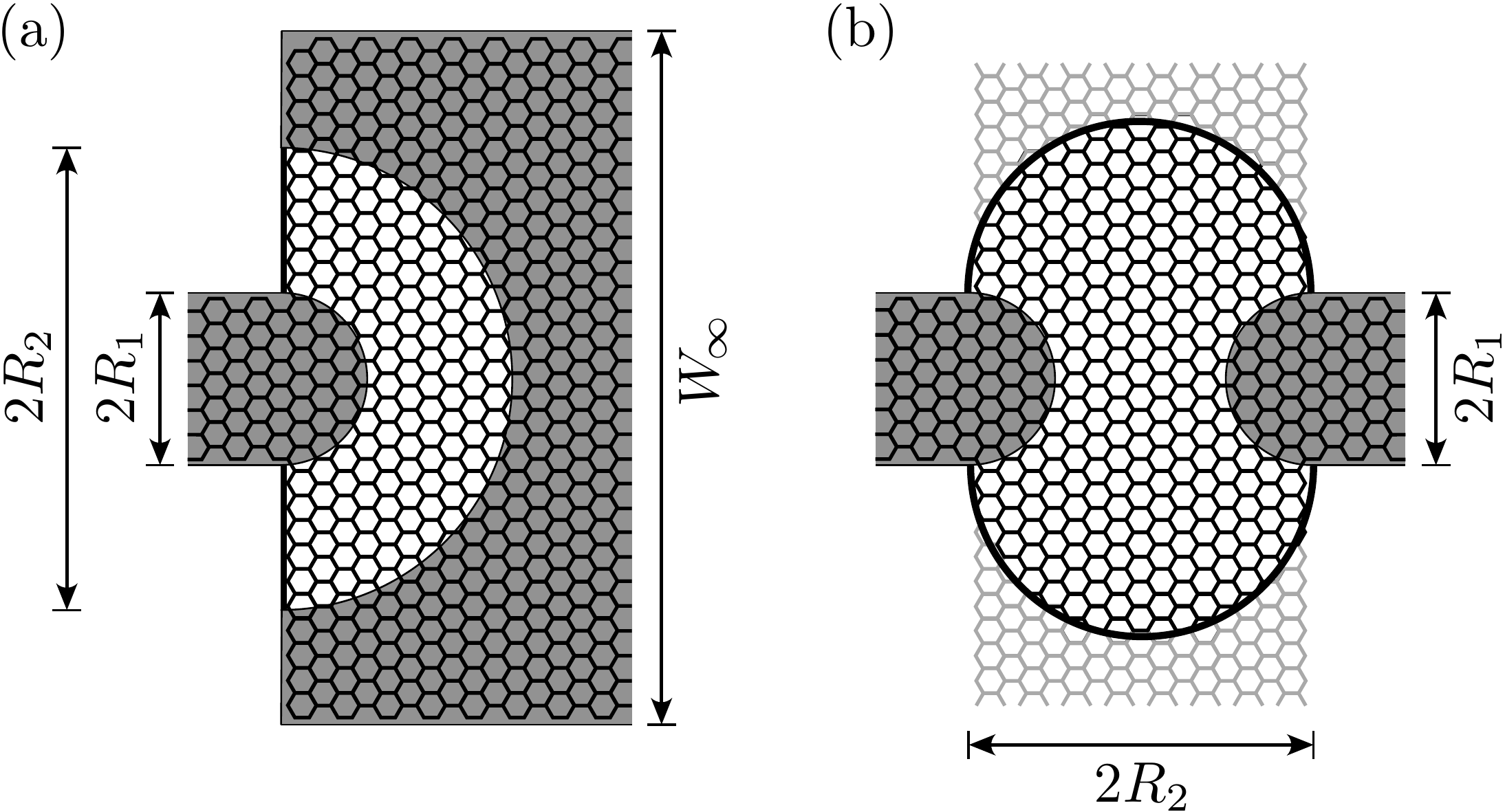}}
\caption{\label{set2term}
  The half-Corbino disk (a) and quantum dot with circular edges (b)
  realized on a honeycomb lattice. Shadow areas mark heavily-doped graphene
  leads. Thick black lines indicate the mass confinement. 
}
\end{figure}

\begin{figure}[!ht]
\centerline{\includegraphics[width=\linewidth]{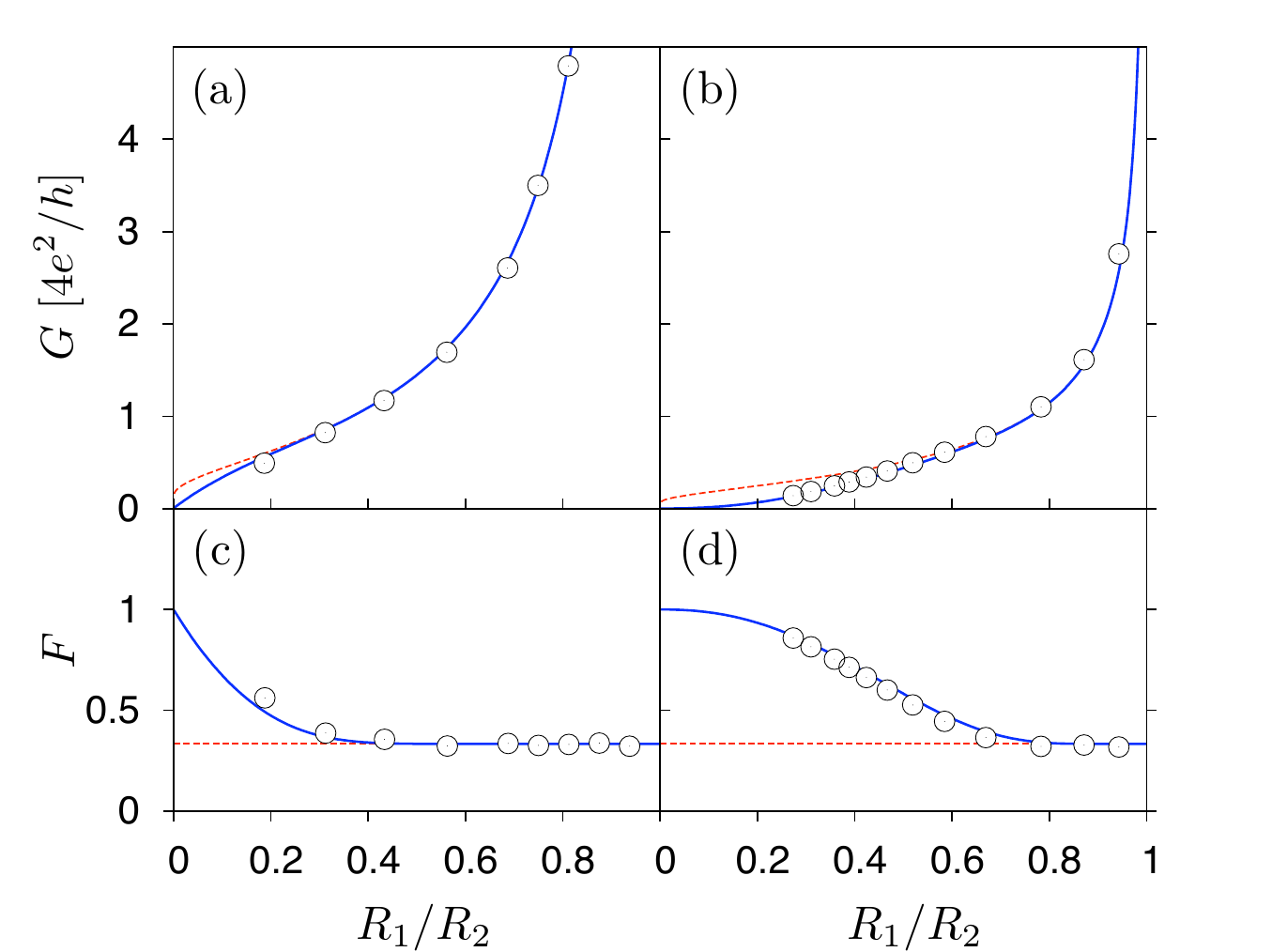}}
\caption{\label{gf2term}
  Conductance (a,b) and Fano factor (c,d) for the half-Corbino disk
  (left) and circular quantum dot (right). 
  Solid lines show the results obtained by numerical summation of 
  (\ref{gsumtj}) and (\ref{fsumtj}) over the modes, dashed lines show the 
  pseudodiffusive limits (\ref{gdifflam}) and (\ref{favertt}). 
  Datapoints on left/right panels are obtained from a computer simulation 
  of transport through the system of Fig.\ \ref{set2term}(a) and 
  \ref{set2term}(b).
}
\end{figure}

We now test the analytical predictions reported earlier in this section by 
comparing them with the results of a computer simulation of electron transport 
in graphene. The discussion starts from the tight-binding model of 
gra\-phe\-ne, with Hamiltonian
\begin{equation}
\label{hami}
  H=\sum_{i,j}\tau_{ij}|i\rangle\langle j|+\sum_{i}V_{i}|i\rangle\langle i|.
\end{equation}
The hopping matrix element $\tau_{ij}=-\tau$ if the orbitals $|i\rangle$ and 
$|j\rangle$ are nearest neighbors on the honeycomb lattice (with 
$\tau=2.7\,\mbox{eV}$), otherwise $\tau_{ij}=0$. The single-particle potential
$V_j$ is arranged such that the chemical potential $\mu_j\equiv E_F-V_j=
\mu_\infty$ in the leads marked by shadow areas in Fig.\ \ref{set2term}, 
whereas between the leads (white area) $\mu_j=0$, except for the small 
regions, where we put $\mu_j=\mu_{A,B}$ (with $\mu_A=-\mu_B$, depending 
whether the atom belong to the $A$ or $B$ sublattice) to model 
a mass-confinement on a honeycomb lattice \cite{Akh08}. Such regions are: 
the outermost edge atoms in the case of the half-Corbino disk (Fig.\ 
\ref{set2term}a), and the atoms placed out of the dot edge (thick lines in 
Fig.\ \ref{set2term}b) for the case of a quantum dot with circular edges.

We have calculated the transmission matrix numerically by adapting the method 
developed by Ando for a square lattice \cite{And91} to the honeycomb lattice. 
The results of our computer simulation \cite{simcocdot}, depicted by 
datapoints in Fig.\ \ref{gf2term}, match theoretical predictions (solid blue 
lines) as long as the number of modes in the narrow lead $N_1\gtrsim 20$. 
Moreover, the formulas (\ref{gcordif}) and (\ref{gcirdif}) for the 
pseudodiffusive conductance (dashed red lines in Fig.\ \ref{gf2term}a,b) 
reproduce the full expression (\ref{gsumtj}) with $1\%$ 
accuracy for $R_1\geqslant 0.29R_2$ in the case of the half-Corbino disk, and 
for $R_1\geqslant 0.69R_2$ in the case of the quantum dot with circular edges. 
Analogously, the pseudodiffusive value of 
the Fano factor $F\approx 1/3$ (see Fig.\ \ref{gf2term}c,d) matches the full
expression (\ref{fsumtj}) with $1\%$ accuracy for $R_1\geqslant 0.43R_2$ and 
$R_1\geqslant 0.81R_2$, respectively. 

In other words, the half-Corbino disk, attached to one narrow and one wide 
lead, represents the case in which electron transport demonstrates the 
pseudodiffusive character in a surprisingly wide range of the system's 
geometrical parameters. On the contrary, in the case of the circular quantum 
dot attached to two narrow leads, both the conductance and the shot noise show 
strong deviations from the pseudodiffusive predictions, as the transport is 
dominated by a single mode in a relatively wide range of parameters. The latter 
represents an example of a graphene system for which our predictions on 
quantum-tunneling transport (such as an approximately \emph{quadratic} decay of
the conductance with $R_2/R_1$) seem to be particularly feasible for an 
experimental verification, also because similar systems have already been 
fabricated \cite{Pon08} suggesting that the role of mass confinement is crucial
when discussing the electronic structure of \emph{closed} quantum dots in 
graphene. Moreover, a recent numerical study shows that the mass confinement 
leads to a strong suppression of weak localization in such systems 
\cite{Wur08}, as observed earlier in experiment \cite{Tik08}.

Below, we extend our numerical analysis to \emph{open}
systems that cannot be obtained from a strip by conformal transformation, 
to illustrate the generic character of the quantum-tunneling transport 
in undoped graphene.

\section{Electron transport \emph{across} a long nanoribbon}
\label{ribb}

\begin{figure}[!ht]
\centerline{\includegraphics[width=\linewidth]{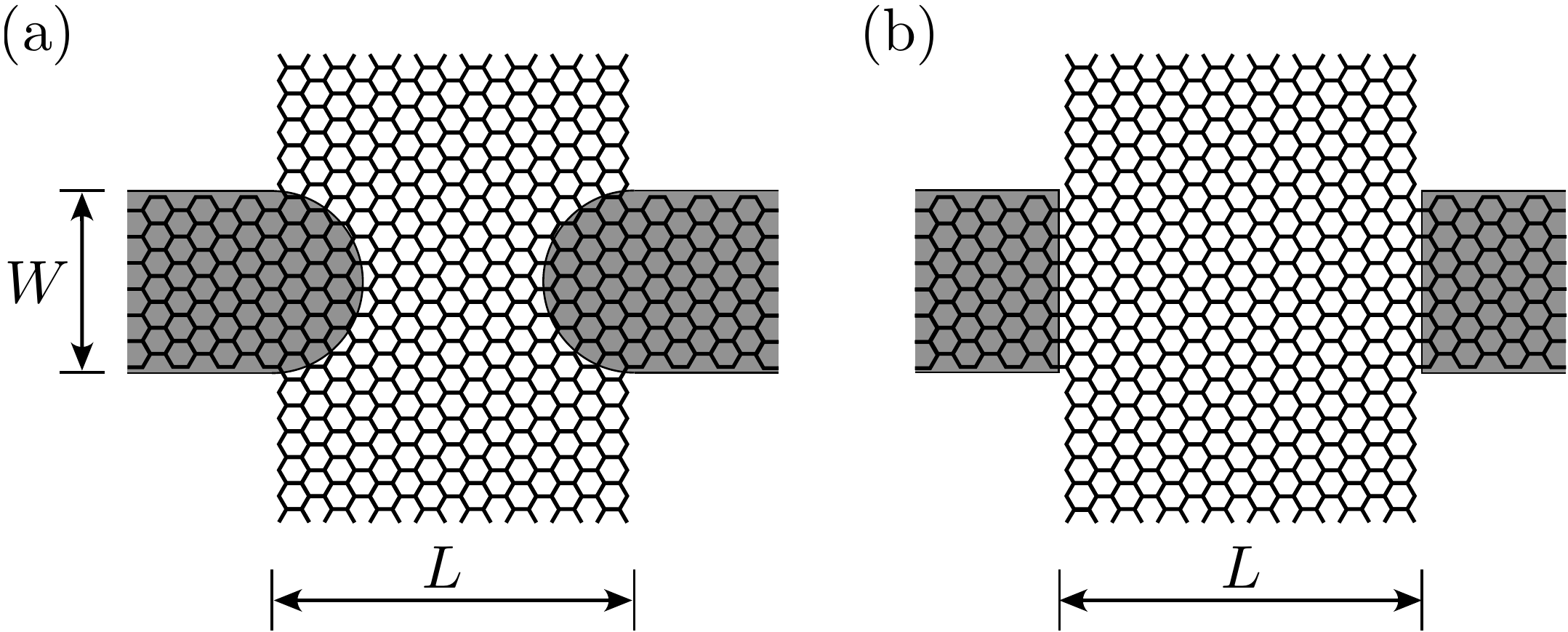}}
\caption{\label{set4term}
  Nanoribbon attached \emph{perpendicularly} to the semicircular (a)
  and rectangular (b) leads. Each system is characterized by the lead width 
  $W$ and the sample area length $L$. Shadow areas mark heavily-doped
  graphene leads.
}
\end{figure}

\begin{figure}[!ht]
\centerline{\includegraphics[width=0.9\linewidth]{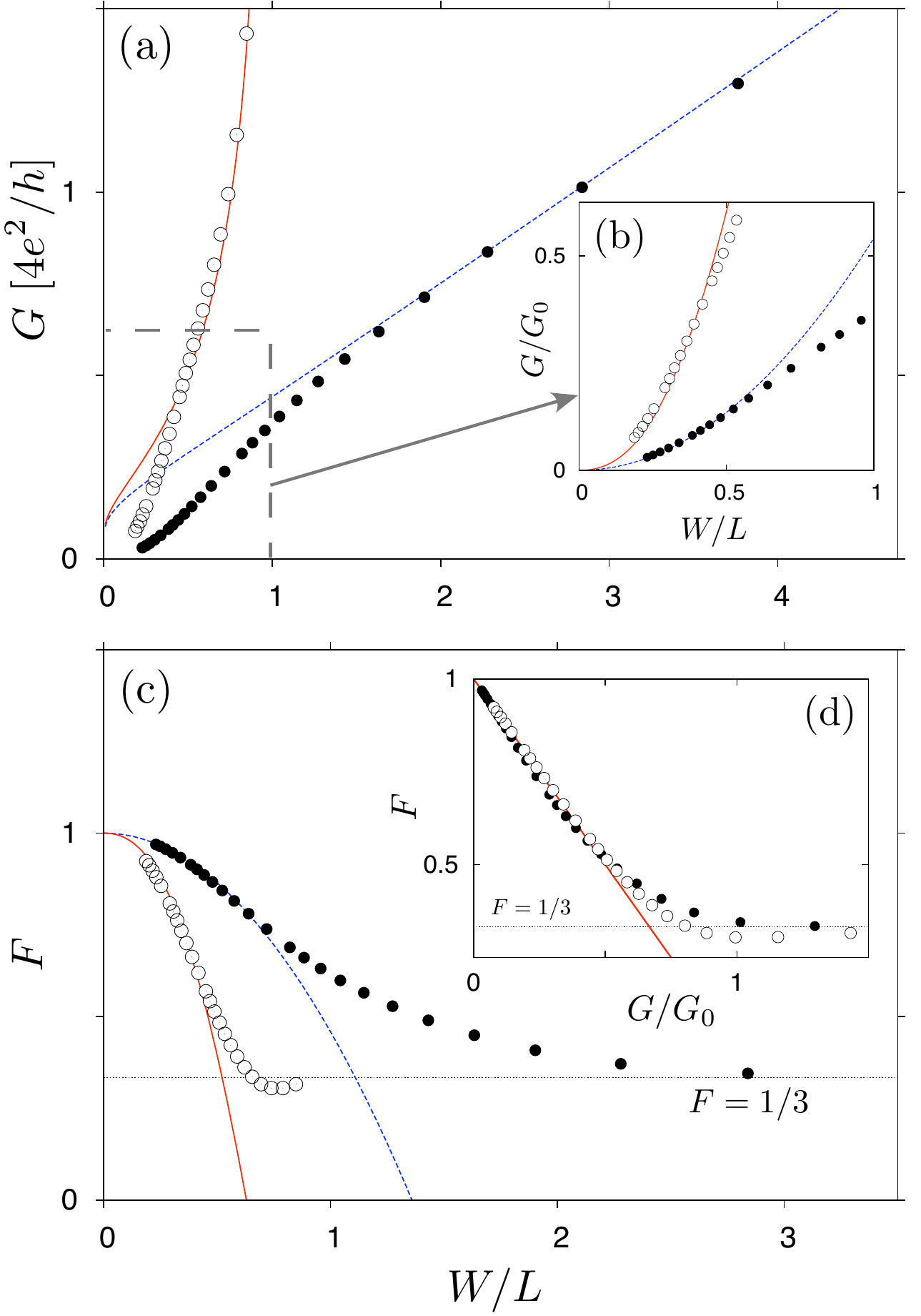}}
\caption{\label{gf4term}
  Conductance (a,b) and Fano factor (c,d) obtained numerically for the system 
  of Fig.\ \ref{set4term}(a) and Fig.\ \ref{set4term}(b) (open and solid 
  symbols in all panels) compared with analytical predictions (lines).
  (a) The pseudodiffusive conductance (\ref{gcirdif}) (solid red line) and 
  (\ref{gsqudif}) (dashed blue line). Solid and dashed lines in panels (b,c): 
  the tunneling conductance (\ref{gciltun}) and (\ref{gsqutun}), and the 
  corresponding values of the Fano factor $F\approx 1-Gh/4e^2$. [The relation
  depicted by the solid line in the shot noise vs conductance diagram (d).] 
  The pseudodiffusive limit $F=1/3$ is shown by the black dotted line (c,d).
}
\end{figure}

In this section, we present
the results obtained from computer simulations of transport \emph{across} 
a long nanoribbon attached to the semicircular 
(Fig.\ \ref{set4term}a) and rectangular (Fig.\ \ref{set4term}b) leads, 
which demonstrate a striking analogy between these systems and the circular
quantum dot studied in the previous section.

Each of the systems in Fig.\ \ref{set4term} is modeled by the
tight-binding Hamiltonian (\ref{hami}). The simulation parameters 
\cite{simribbon} are chosen to grasp the basic features of recently fabricated 
graphene nanoribbons \cite{Li08}, which have zigzag edges and are insulating, 
as the weak staggered potential placed at the ribbon edge opens a band gap in 
the electronic spectrum \cite{Son06}. A similar effect was observed in
recent numerical studies of long nanoribbons with weak edge disorder 
\cite{Muc09} or irregular edges \cite{Rai09}.

\subsection{Results for an \emph{infinitely} long ribbon}
We utilize the $4$-terminal recursive Green's function algorithm 
\cite{Wim08a}, which allows us to analyze directly the electron transport 
across an \emph{infinitely} long nanoribbon in graphene. Namely, we attached 
two extra leads (one from the top and one from the bottom, not shown)
to each of the systems in Fig.\ \ref{set4term}, that are undoped and thus
contain the evanescent modes only. (Notice that the chemical potential for the 
outermost edge-atoms $\mu_{A,B}\neq 0$ \cite{simribbon}.) 
The results are shown in Fig.\ \ref{gf4term}.

\begin{figure*}[!t]
\centerline{\includegraphics[width=0.8\linewidth]{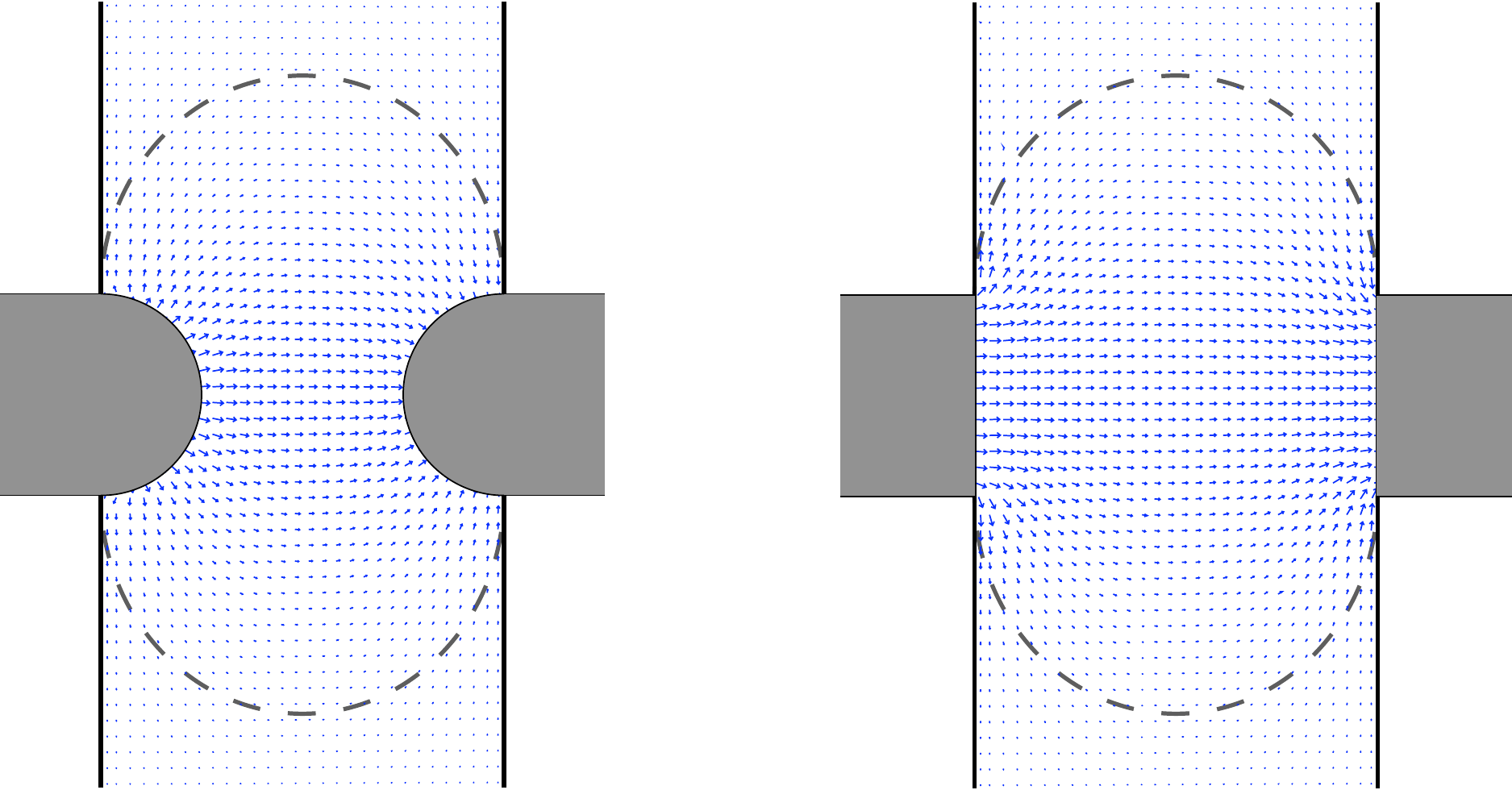}}
\caption{\label{jxy4te}
  Current distribution in a long nanoribbon attached to circular (left) and 
  rectangular (right) leads, as shown in Fig.\ \ref{set4term}(a,b). Each arrow 
  represents the average current density for a~rectangle consisting of 
  $17\times 17$ unit cells. The aspect ratio of both systems is 
  $W/L\approx 0.5$. Dashed lines mark symbolically the edges of 
  the corresponding quantum dot of Fig.\ \ref{set2term}(b).
}
\end{figure*}

The conductance of a nanoribbon attached perpendicularly to circular
leads (top panel in Fig.\ \ref{gf4term}, open symbols) approaches the 
asymptotic formula for the circular quantum dot (\ref{gcirdif}) with $R_1=W/2$,
$R_2=L/2$, $\vartheta=\pi$ for $W\approx L$ (solid red line). For instance, 
a $2\%$ agreement is reached at $W/L=0.85$. This is a consequence of the fact 
that in the absence of propagating modes in a ribbon, most of the current 
flows via the central region of the device, and the system of 
Fig.\ \ref{set4term}a becomes effectively identical to the circular quantum 
dot in the pseudodiffusive limit, where the role of boundary conditions is 
negligible. For the opposite, quantum-tunneling limit $W\ll L$ the 
corresponding formula (\ref{gcirtun}) may be written as
\begin{equation}
\label{gciltun}
  G\approx 2^{\eta}\pi\sigma_0\left(W/L\right)^{2-\eta},\ \ \ \ \ \ 
  \mbox{with}\ \ \ \eta\equiv 4W/\pi L,
\end{equation}
what agrees surprisingly well with the actual data shown in Fig.\ \ref{gf4term}
(see inset in the top panel; solid red line and open symbols, respectively).
Such an agreement can be understood when looking at the current-density
distribution, shown in Fig.\ \ref{jxy4te}. Even for an aspect ratio as small
as $W/L\approx 0.5$, over $90\%$ of the current does not leave the area of 
a circular quantum dot (bounded symbolically with dashed lines).

For the case of a nanoribbon attached perpendicularly to rectangular leads 
(Fig.\ \ref{set4term}b), the pseudo-diffusive conductance (for $W\gg L$) is 
given by \cite{pdifsqu}
\begin{equation}
\label{gsqudif}
  G_\mathrm{diff}=\frac{\sigma_0W}{2L}\frac{\pi}{\arctan\left(\frac{W}{L}\right)
  +\left(\frac{W}{L}\right)\ln\sqrt{1+\left(\frac{L}{W}\right)^2}},
\end{equation}
which is depicted in top panel of Fig.\ \ref{gf4term} (dashed blue line) and 
matches the numerical data (solid symbols) within $2\%$ accuracy for $W/L
\gtrsim 2$. An identically good agreement with the numerics is observed for the 
asymptotic form of the formula (\ref{gsqudif}) $G_\mathrm{diff}\approx\sigma_0
\left(W/L+1/\pi\right)$, showing that the infinite ribbon attached 
perpendicularly to the leads has an extra $\sigma_0/\pi$ conductance in 
comparison with the rectangular geometry considered in Refs.\ 
\cite{Two06,Ryc08,Mia07,Dan08}.

A brief comparison between the formula (\ref{gsqudif}) and the generic form
of the pseudodiffusive conductance (\ref{gdifflam}) allows us to consider 
the functional $\Lambda\equiv\Lambda(W/L)$ in an approximate form given by
\begin{equation}
  \ln\Lambda\approx 2\left[\ln\left(\frac{W}{L}\right) + 
    \left(\frac{L}{W}\right)\arctan\left(\frac{W}{L}\right)\right].
\end{equation}
Subsequently, an approximate form of the quantum-tunneling conductance 
(\ref{gftunnlam}) for $W\ll L$ is
\begin{equation}
\label{gsqutun}
  G\approx 4\pi\sigma_0\Lambda^{-1}\approx 4\pi\sigma_0e^{-2}\left(W/L\right)^2.
\end{equation}
Again, the formula (\ref{gsqutun}) shows a surprisingly good, approximately
$10\%$ agreement with the numerical data presented in Fig.\ \ref{gf4term} (see 
the inset in the top panel; dashed blue line and solid symbols, respectively),
suggesting that the power-law (approximately quadratic) decay of $G$ for large 
$L$ is a generic feature for transport across the nanoribbon, unrelated to the
particular shape of the leads \cite{Katfoo}.

The numerical results for the shot-noise power are presented in the bottom 
panel of Fig.\ \ref{gf4term}. The approximative formulas (\ref{gciltun}) and 
(\ref{gsqutun}) are substituted in the relation $F\approx 1-Gh/(4e^2)$, which
produces the analytical predictions depicted by solid red and dashed blue
lines, respectively. In both cases, the agreement with numerical results is 
better than $5\%$ when the Fano factor $F\gtrsim 2/3$. 
An additional insight into the nature of the crossover from the Poissonian to 
the pseudodiffusive regime is provided by an $F$ versus $G$ plot (see the 
inset). In particular, values of $F$ are very close 
to $1-Gh/(4e^2)$ even for a relatively large conductance $G\approx 2e^2/h$,
which indicates that electron transport is governed by a single, 
valley-degenerated mode in a wide 
range of the geometrical parameters ($W/L\lesssim 0.5$ for the circular leads, 
and $W/L\lesssim 1.5$ for the rectangular leads).

\begin{figure}[!ht]
\centerline{\includegraphics[width=0.8\linewidth]{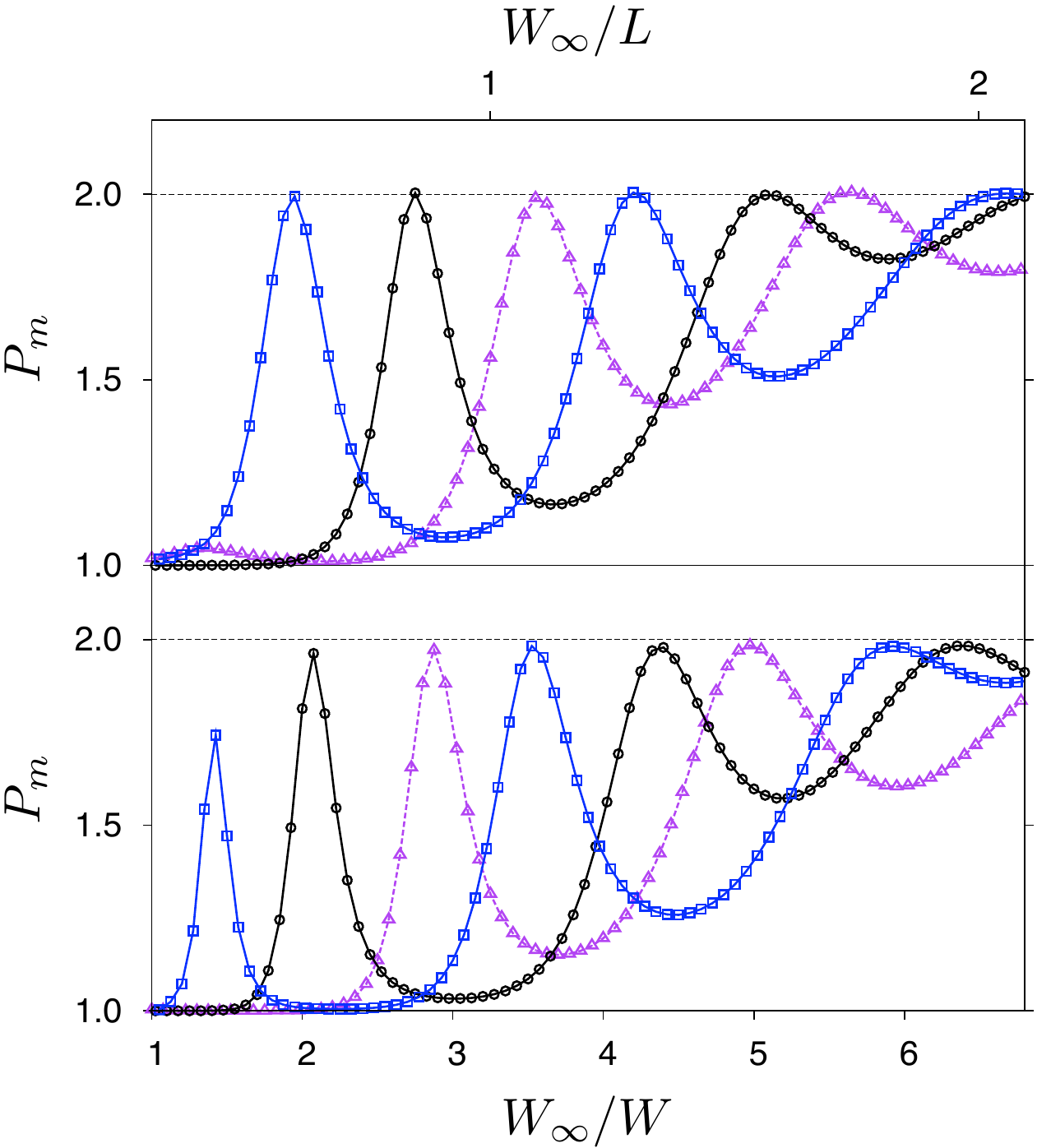}}
\caption{\label{pmod}
  Mode-participation ratio for transport through the system of Fig.\ 
  \ref{set4term} with $W/L\approx 0.3$ and finite width of the undoped region 
  $W_\infty$. Three curves on each panel correspond to $W_\infty/a=3k$ (squares),
  $3k+1$ (circles), and $3k+2$ (triangles), with $k$-integer. 
  Top: circular leads, bottom: rectangular leads. 
  Lines are drawn as a guide for the eye only.
}
\end{figure}

\subsection{Influence of armchair edges in a \emph{finite} ribbon}
So far, we have analyzed the transport across an \emph{infinitely} long, 
zigzag nanoribbon attached perpendicularly to the leads. To find out how 
the results change for the realistic case of a long but \emph{finite} 
nanoribbon, we consider now the system of Fig.\ \ref{set4term} with a central 
(undoped) region of finite width $W_\infty$. The system is terminated from the 
top and the bottom by armchair boundaries which mix valley degrees of freedom
\cite{Two06}, so the fourfold (spin and valley) degeneracy of transmission
eigenvalues $T_j$ is expected to be replaced by the twofold (spin only) 
degeneracy. To trace the effect of armchair boundaries in a quantitative
manner, we define the mode-participation ratio
\begin{equation}
\label{pmoddef}
  P_m=\frac{\left(\sum_jT_j\right)^2}{\sum_jT_j^2}=
  \frac{2}{1-F}\frac{G}{\pi\sigma_0},  
\end{equation}
where we assume spin-only degeneracy in summations. 
In particular, for the quantum-tunneling limit $W\ll L$, 
the mode-participation ratio is $P_m\approx 2$ if the lowest mode,
that governs the electronic transport, has an approximate valley degeneracy. 
Otherwise, in this limit $P_m\approx 1$.

The numerical values of the mode-participation ratio (\ref{pmoddef}) are
presented in Fig.\ \ref{pmod}. We took $W=80a$ (providing $30$ propagating 
modes for $\mu_\infty=\tau/2$), $L=150\sqrt{3}a$ (so $W/L\approx 0.3$) and vary 
$W_\infty$. The remaining parameters are identical as in the case of an 
infinite ribbon, studied before. The datapoints in Fig.\ \ref{pmod} 
illustrate a smooth crossover from the transport dominated by a single mode 
with spin-only degeneracy ($W_\infty\sim W\ll L$), to the situation with full
fourfold degeneracy ($W\ll L\ll W_\infty$). The details of the evolution depends
on whether the width $W_\infty$ corresponds to the metallic ($W_\infty/a=3k+1$) 
or to one of the two insulating armchair boundary conditions ($W_\infty/a=3k$, 
$3k+2$). In all cases, the valley degeneracy is approximately restored 
($P_m\approx 2$, marked with thin black line) for $W_\infty\gtrsim 2L$, when 
the role of armchair edges becomes negligible, as the current is flowing 
predominantly via the central area of the system (see the current distribution
shown in Fig.\ \ref{jxy4te}, right panel).

\subsection{Implications for the experiment}
\begin{figure}[!ht]
\centerline{\includegraphics[width=0.8\linewidth]{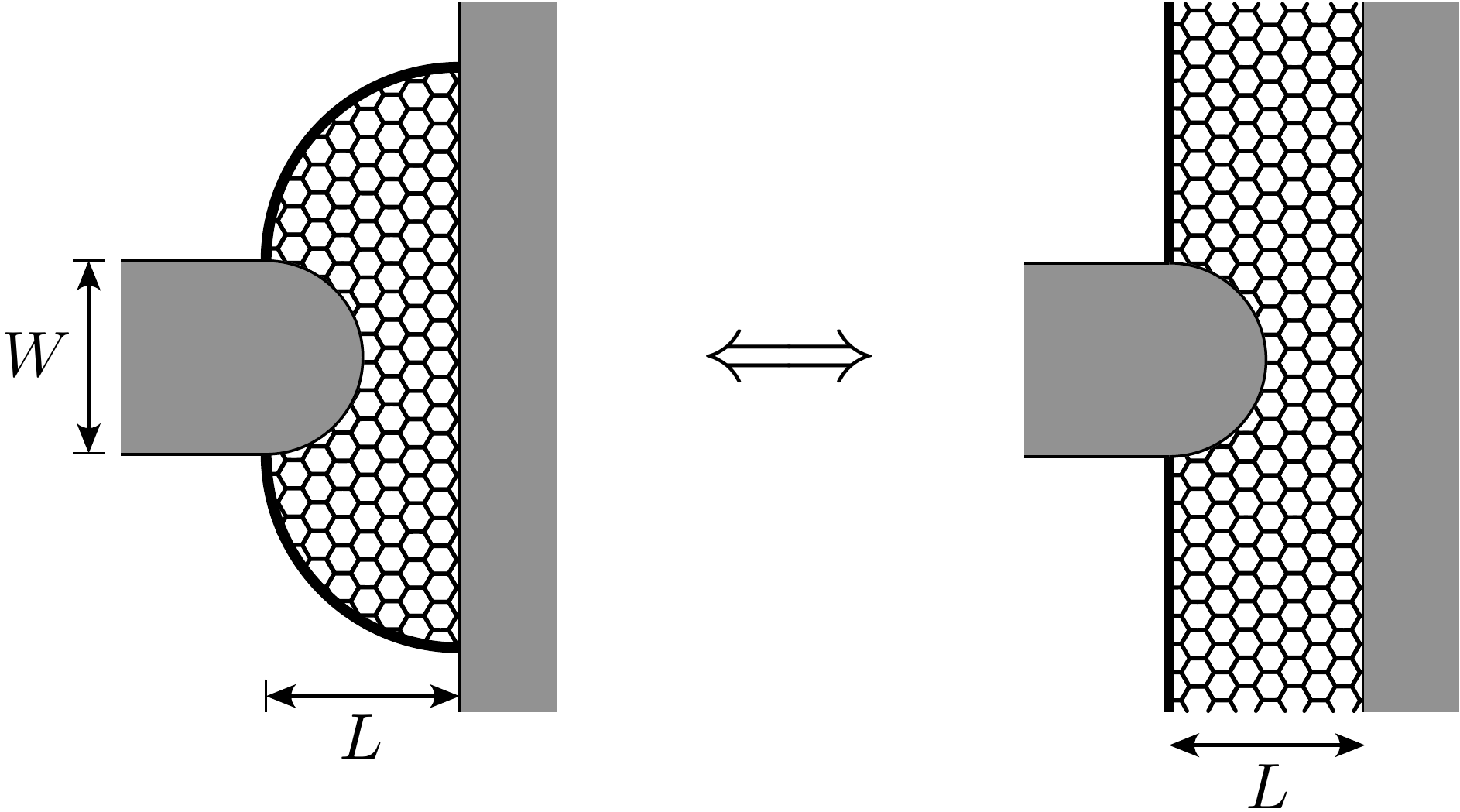}}
\caption{\label{cirsqu}
  These two graphene billiards both have the same conductance and shot noise
  in the pseudodiffusive regime $W\lesssim 2L$.
}
\end{figure}

For the sake of completeness, we analyze now the transport through a graphene 
billiard attached to two different leads, one narrow and semicircular, and the 
other wide and rectangular, as shown in Fig.\ \ref{cirsqu}. As before,
mass confinement (thick black lines) is applied for the edges not 
connected to the leads (shadow areas). The system shown in the left panel
of Fig.\ \ref{cirsqu} can be exactly mapped onto a strip (see Fig.\ 
\ref{interf12}) by the conformal transformation (\ref{contra2}) with the 
condition $(l'+2L-W)/(l'-2L+W)=e^{\vartheta_\infty\mathcal{L}/\mathcal{W}}$, where
$l'\equiv\sqrt{4L^2-W^2}$. This implies the functional
\begin{equation}
\label{lacontra2p}
  \Lambda(W/L) = \left(\frac{l'+2L-W}{l'-2L+W}\right)^{\pi/\vartheta_\infty},
\end{equation}
where $\vartheta_\infty=2\pi-2\arcsin[l'/(2L)]$ is the angle with which edges 
intersect each other in the poles of conformal transformation (see Fig.\ 
\ref{corbcido}c, with $\vartheta=\pi$). The pseudodiffusive (\ref{gdifflam}) 
and the quantum-tunneling conductance now take on the forms
\begin{equation}
\label{gdiffcisq}
  G_\mathrm{diff}=\frac{\sigma_0\vartheta_\infty}{\ln
  \left[(l'+2L-W)/(l'-2L+W)\right]},
\end{equation}
with $\vartheta_\infty\rightarrow 2\pi$ for $W/L\rightarrow 1$, and
\begin{equation}
\label{gtunncisq}
  G\approx 4\pi\sigma_0\left(\frac{l'-2L+W}{l'+2L-W}\right)^{\pi/\vartheta_\infty}
  \approx \pi\sigma_0\frac{W}{L},
\end{equation}
where the last asymptotic expression refers to the limit $W/L\rightarrow 0$,
for which the system behaves effectively like the half-Corbino disk with
the inner radius $R_1=W/2$ and the outer radius $R_2=2L$. As a consequence, 
the pseudodiffusive conductance (\ref{gdiffcisq}) reproduces the values
obtained from the exact expression \cite{gcisq} with 1\% accuracy for
$W/L\geqslant 0.82$.

\begin{figure}[!ht]
\centerline{\includegraphics[width=0.8\linewidth]{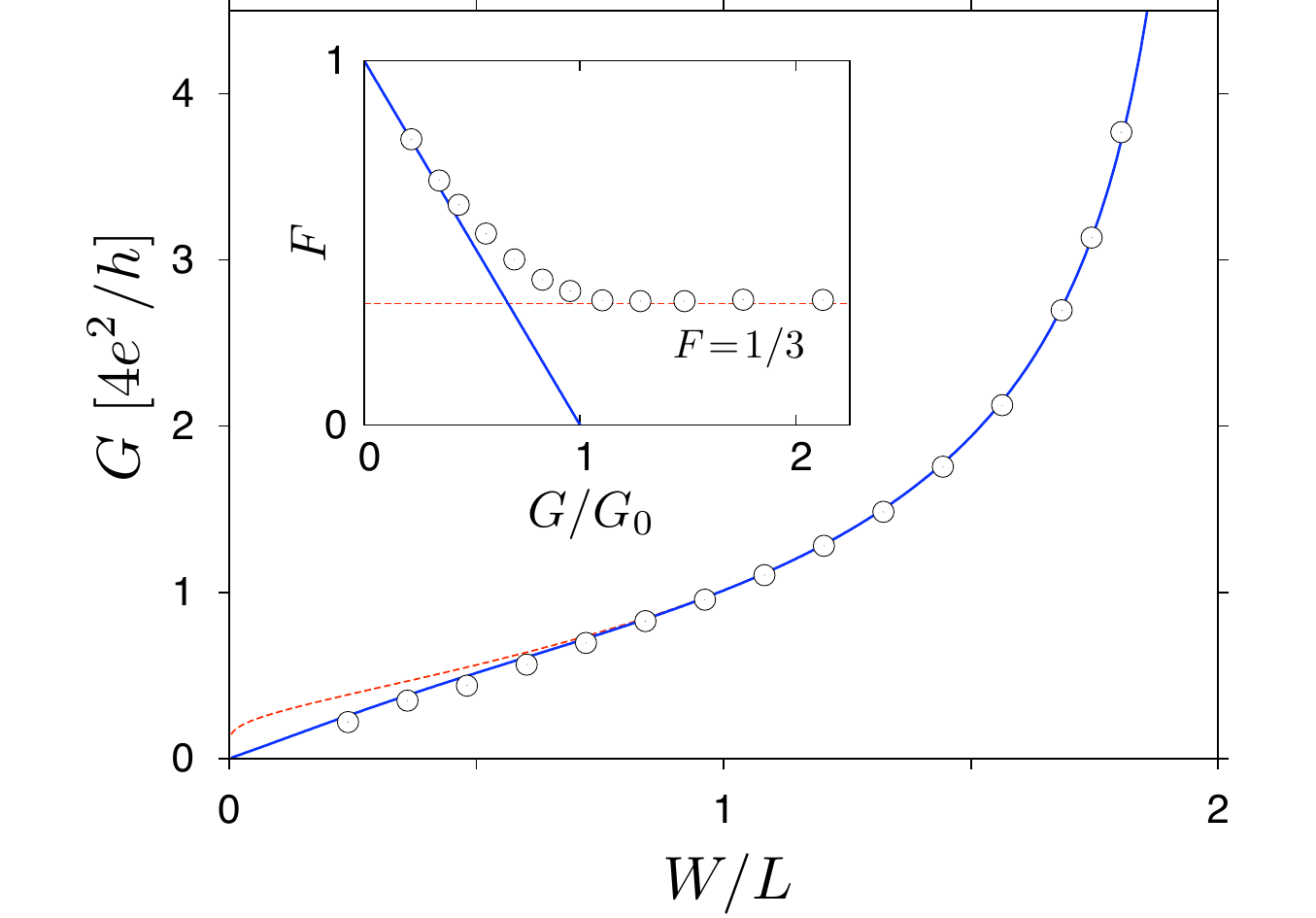}}
\caption{\label{gfcisq}
  Conductance and shot noise of the two systems shown in Fig.\ \ref{cirsqu}.
  Main panel: The exact (solid line) and the pseudodiffusive (dashed line) value
  of conductance for a billiard with circular edges, and the results of 
  computer simulation for a nanoribbon-like billiard (datapoints). Inset: Shot 
  noise vs conductance diagram for a nanoribbon-like billiard. Solid and dashed
  lines mark the asymptotic values for the tunneling and the pseudodiffusive 
  limit, respectively.
}
\end{figure}

Another striking feature of the pseudodiffusive regime $W\lesssim 2L$, 
which coincides with findings presented earlier in this section, is related
to the fact that poles of the conformal transformation (\ref{contra2})
approach the circular lead tip for $W/L\rightarrow 2$. This is why the current
is flowing mainly through the central area of the system, and the two billiards
shown in Fig.\ \ref{cirsqu} become equivalent in such a limit. The earlier
findings for the circular quantum dot and the long nanoribbon attached 
perpendicularly to the leads, allow us to expect that the conductance of the
nanoribbon-like system shown in the \emph{right} panel of Fig.\ \ref{cirsqu}
will not deviate significantly from the expression \cite{gcisq}
also in the tunneling limit $W\ll L$. The numerical results presented in Fig.\ 
\ref{gfcisq} confirm such an expectation, as the conductance obtained by a
computer simulation for the nanoribbon-like system \cite{simcisq} match again
the analytical predictions for the system with circular edges in surprisingly
wide range of the parameters. Namely, $10\%$ agreement is reached for 
$W/L\geqslant 0.4$, whereas for $W/L\geqslant 0.8$ the deviation drops below 
$2\%$.

We predict that the pseudodiffusive conductance (\ref{gdiffcisq}) remains 
unchanged for a wide class of irregular graphene billiards of shapes fitting 
\emph{between} the two limiting cases shown in Fig.\ \ref{cirsqu}. Moreover, 
an approximate agreement should be observed even for the tunneling conductance 
(\ref{gtunncisq}). We believe that such an extra flexibility in device setups
will facilitate experiments with better agreement with the theory as
achieved so far for rectangular samples \cite{Mia07,Dan08}. In particular,
the setup consisting of the narrow semicircular lead on one side and the 
straight graphene-lead interface on the other side, eliminates the difficulty
of manufacturing the two \emph{parallel} interfaces---one of the main
problems that have limited the number of experimental samples, suitable for 
both ballistic conductance and shot-noise measurements, to just a~few so far.

\section{Conclusions}
In conclusion, we have identified a novel type of quantum tunneling effect,
which appears in transport through the Corbino disk and quantum billiards in 
undoped graphene, provided that at least one of the leads (or billiard 
openings) is much narrower than the distance between openings $L$, which 
defines the length-scale of the sample. In such a tunneling limit, the 
conductance $G$ shows a slow power-law decay with $L$ characterized by a 
geometry-dependent exponent. The Fano factor $F$ exhibits a crossover from the 
pseudodiffusive ($F=1/3$) to Poissonian ($F=1$) shot noise, with a relation 
$G\approx(1\!-\!F){\times}se^2/h$ in a~surprisingly wide range of $L$. 
This is because electron transport in the tunneling limit is effectively 
governed by a single mode, having the full spin, valley, and symplectic 
degeneracy ($s=8$) in the absence of boundaries (Corbino geometry), spin and 
valley degeneracy ($s=4$) if the boundary conditions do not scatter valleys, 
or the spin-only degeneracy ($s=2$) otherwise. In particular, for the case of 
a ribbon which contains either infinite-mass or armchair boundaries, the valley
degeneracy is restored when armchair endings are shifted away from the area 
where the main current flows. We would like to stress that the relation between
$G$ and $F$ allows an experimental verification of the degeneracy $s$ without 
referring to any geometrical parameters.

We have explored the idea of Katsnelson and Guinea \cite{Kat08}, that 
transmission eigenvalues could be obtained analytically for any undoped 
graphene flake of a geometry linked via conformal transformation to a strip, 
for which the solution is known due to Tworzyd{\l}o et al.\ \cite{Two06}. 
In the pseudodiffusive limit, we show that the eigenvalue distribution is 
affected by an arbitrary conformal transformation only via a multiplicative
prefactor, and the value $F=1/3$ is unchanged for any \emph{closed} setup, 
provided that $G\gg e^2/h$. We test the approach for the Corbino disk, by 
comparing transmission probabilities obtained by a conformal mapping, and 
within the mode-matching analysis for angular momentum eigenstates.
To analyze a crossover from the tunneling to the pseudodiffusive limit in a
confined system, we focus on two particular billiards (a section of the 
Corbino disk and the quantum dot with circular edges) confined by a mass. 
The results of our numerical simulation of transport through a~lattice 
consisting of ca.\ $10^5$ carbon atoms match the expressions for $G$ and $F$ 
obtained by conformal mapping. 
Moreover, we generalize the approach to obtain an approximate
formula for $G$ (which reproduces either the pseudodiffusive values or the 
tunneling-limit exponent resulting from the simulation) in the case of an 
infinite ribbon attached perpendicularly to the leads---an \emph{open} 
billiard, not linked to the strip via conformal transformation.

\section*{Acknowledgment}
We thank Bj\"{o}rn Trauzettel for helpful discussions.
A.R.\ acknowledges the support from the Alexander von Humboldt Stiftung-Foundation, the Polish Ministry of Science (Grant No.\ N--N202--128736), and the Polish Science Foundation (FNP). We further acknowledge financial support from the German Research Foundation (DFG) via Tr950/1--1  (P.R.) and SFB~689 (M.W.).

\appendix
\section{Conformal mapping for a generic setup with circular contacts}
\label{appcir}

\begin{figure}[!ht]
\centerline{\includegraphics[width=0.8\linewidth]{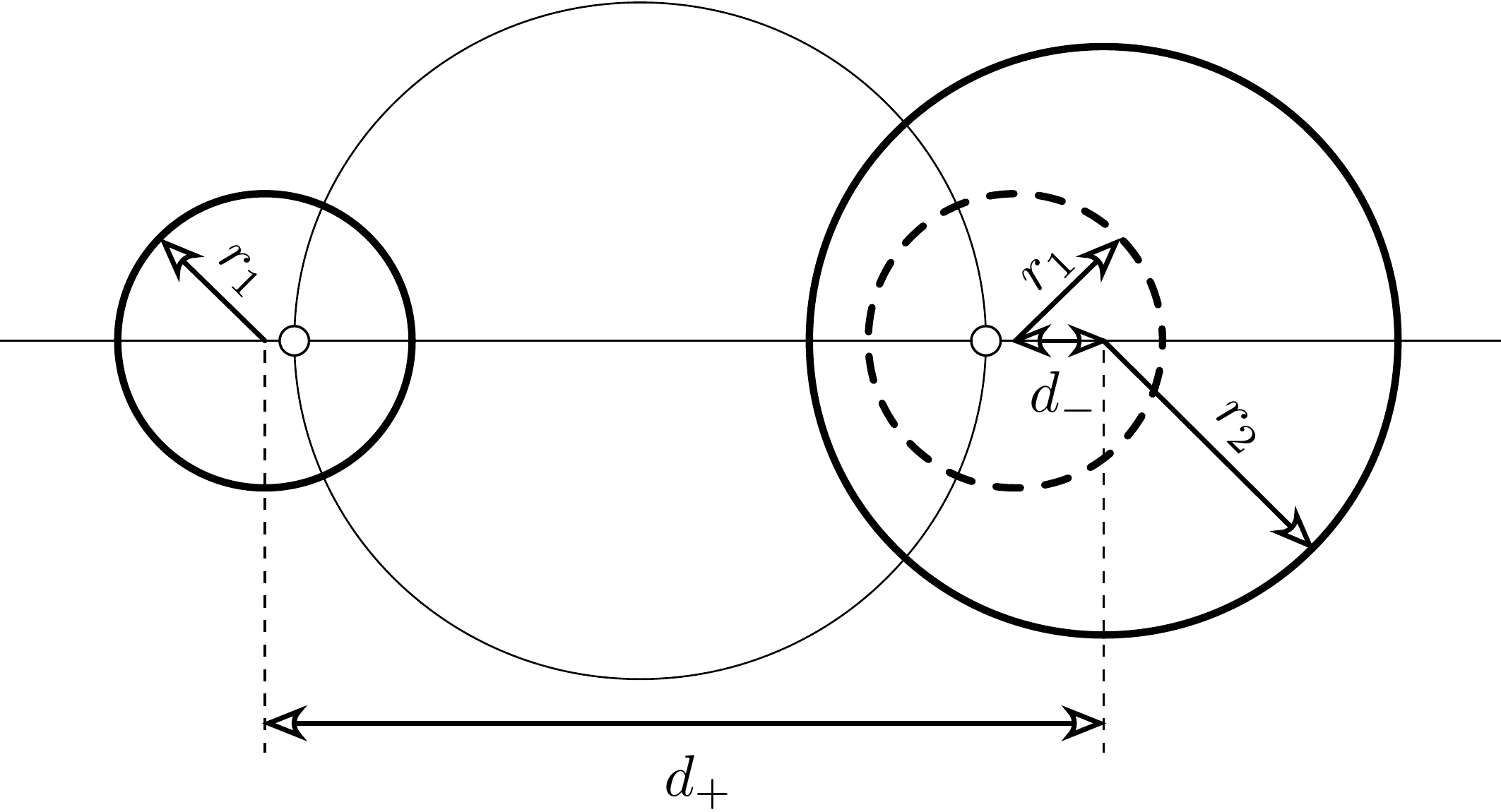}}
\caption{\label{nocoacir}
  Generic circular contacts (thick lines) of radii $r_1$ and $r_2>r_1$ 
  misplaced by a distance $d_+>r_1+r_2$ or $d_-<r_2-r_1$ (dashed line depicts 
  the \emph{inner} contact interface for the latter case). White dots mark 
  poles of the transformation (\ref{mobius}), thin lines are perpendicular to 
  each of the interfaces (and mapped onto the radiant lines by the 
  transformation). 
}
\end{figure}

We consider here a generic setup, containing two circular, but not coaxial 
interfaces splitting undoped and heavily doped graphene interfaces, as depicted
schematically in Fig.\ \ref{nocoacir}. In the first case, an infinite graphene
plane is probed by the leads of radii $r_1$ and $r_2$ (thick solid lines), 
misplaced by the distance $d_+>r_1+r_2$ (we further suppose $r_1<r_2$) and 
heavily doped. In the second case, a disk-like sample area is 
limited by the inner lead of radius $r_1$ (dashed circle) and the outer lead 
of radius $r_2$, misplaced by $d_-<r_2-r_1$ (and for $d_-\rightarrow 0$ the 
perfect Corbino geometry is restored). In both situations, conformal mapping 
onto the Corbino disk is provided by the M\"{o}bius transformation
\begin{equation}
\label{mobius}
  z=\frac{w-\beta}{w+\beta},
\end{equation}
where $z$ belongs to the disk area (with the edges radii $R_1$ and $R_2$, 
$R_1\leqslant|z|\leqslant R_2$) and $w$ belongs to the sample area of Fig.\ 
\ref{nocoacir}. The real parameter $\beta$ is adjusted such that
\begin{equation}
  d_\pm=\sqrt{r_2^2+\beta^2}\pm\sqrt{r_1^2+\beta^2},
\end{equation}
which leads to the useful relation 
\begin{equation}
\label{ddr12}
  d_+d_-=r_2^2-r_1^2.
\end{equation} 
(Notice that we are using one form of $z(w)$ to describe the two 
distinct situations, in each of which only one of the displacements 
$\{d_+,d_-\}$ is a physical parameter.)

The explicit form of the
functional $\Lambda\equiv (R_2/R_1)^{1/2}$ follows from the condition that
the transformation (\ref{mobius}) always maps the first contact (of radius 
$r_1$) onto the inner edge of the disk, whereas the second contact is mapped
onto the outer edge. For the two situations studied here 
\begin{equation}
\label{lamdpm}
  \Lambda(r_1,r_2,d_\pm)=
  \left(\frac{x_1+y_1}{x_1-y_1}\,\frac{x_2\pm y_2}{x_2\mp y_2}\right)^{1/2},
\end{equation}
where we define the variables $x_\alpha=d_++s_\alpha d_-+2r_\alpha$, 
$y_\alpha=\sqrt{(d_++s_\alpha d_-)^2-\left(2r_\alpha\right)^2}$
(with $s_\alpha\equiv 2\alpha-3$, $\alpha=1,2$).
The remaining one of the parameters $\{d_+,d_-\}$, not listed explicitly
as an argument of $\Lambda$, is determined by Eq.\ (\ref{ddr12}).

In particular, for the case of two identical circular leads probing a large 
graphene plane (see Fig.\ \ref{flak2prob}a) $r_1=r_2\equiv r$, $d_-=0$, and 
the functional
\begin{equation}
  \Lambda(r,d_+)=\frac{d_++2r+\sqrt{d_+^2-4r^2}}{d_++2r-\sqrt{d_+^2-4r^2}}
  \approx \frac{d_+}{r},
\end{equation}
where the approximation refers to the $r\ll d_+$ limit.
Defining $l\equiv d_+$ we obtain Eq.\ (\ref{gtunnflak2a}) of the main text. 
Similarly, taking the limit $r_2,d_+\rightarrow\infty$ such that 
$l'\equiv d_+-r_1-r_2=\mathrm{const}$, we find from Eq.\ (\ref{lamdpm}) that
$\Lambda\approx (2l'/r)^{1/2}$ for $r\equiv r_1\ll l'$, what leads to
Eq.\ (\ref{gtunnflak2b}) for the conductance.

\section{Mode-matching for the Corbino disk}
\label{appmod}

Here, we derive the transmission and reflection amplitudes for scattering
eigenstates of the Hamiltonian (\ref{dirhamcir}) for the Corbino setup, as
shown in Fig.\ \ref{corbcido}(a). Without loss of generality, we suppose 
electron doping in the leads $E>U_\infty$, but an arbitrary doping in the 
sample area $\eta\equiv\mbox{sgn}(E-U_0)=\pm 1$. 

The radial component $\chi_j(r)$ of the eigenstate $\psi_j$ 
(\ref{jeigen}) corresponding to the total angular momentum $\hbar j$ ($j$ 
half-odd integer) and energy $E$ can be divided into three regions.
For $r>R_{2}$ (the outer lead), $\chi_{j}\equiv\chi_j^\mathrm{I}$, with
\begin{equation}
\chi_{j}^\mathrm{I}=\left(\begin{array}{c}H_{j-1/2}^{(2)}(Kr)\\iH_{j+1/2}^{(2)}(Kr)\end{array}\right)+r_{j}\,\left(\begin{array}{c}H_{j-1/2}^{(1)}(Kr)\\iH_{j+1/2}^{(1)}(Kr)\end{array}\right),
\end{equation}
where $K=|E-U_\infty|/\hbar v_F$ with $U_{\infty}\rightarrow-\infty$ \cite{asshan}, and $r_{j}$ the reflection coefficient. Next, for $R_{1}<r<R_{2}$ (the disk area), $\chi_{j}\equiv\chi_j^\mathrm{II}$, with
\begin{equation}
\chi_{j}^\mathrm{II}=a_j\,\left(\begin{array}{c}H_{j-1/2}^{(2)}(kr)\\i{\eta}H_{j+1/2}^{(2)}(kr)\end{array}\right)+b_j\,\left(\begin{array}{c}H_{j-1/2}^{(1)}(kr)\\i{\eta}H_{j+1/2}^{(1)}(kr)\end{array}\right),
\end{equation}
where $k=|E-U_0|/\hbar v_F$. Finally, for $r<R_{1}$ (the inner lead), $\chi_{j}\equiv\chi_j^\mathrm{III}$, with
\begin{equation}
\chi_{j}^\mathrm{III}=t_{j}\left(\begin{array}{c}H_{j-1/2}^{(2)}(Kr)\\iH_{j+1/2}^{(2)}(Kr)\end{array}\right).
\end{equation}\vspace{0.5em}\\
Solving the matching conditions $\chi_j^\mathrm{I}(R_2)=\chi_j^\mathrm{II}(R_2)$ and $\chi_j^\mathrm{II}(R_1)=\chi_j^\mathrm{III}(R_1)$  we find
\begin{multline}
a_j=\sqrt{\frac{2}{\pi K R_{2}}}\,e^{-i\kappa_{2}}\left(\eta\mathfrak{D}_{j}^-+i\mathfrak{D}_{j}^+\right)^{-1}\\ \times\left[H_{j-1/2}^{(1)}(\rho_1)+i{\eta}H_{j+1/2}^{(1)}(\rho_1)\right],
\end{multline}
\begin{multline}
b_j=-a_j\,\frac{H_{j-1/2}^{(2)}(\rho_1)+i{\eta}H_{j+1/2}^{(2)}(\rho_1)}{H_{j-1/2}^{(1)}(\rho_1)+i{\eta}H_{j+1/2}^{(1)}(\rho_1)},\hfill
\end{multline}
where $\rho_{\alpha}=kR_{\alpha}$ (with $\alpha=1,2$), $\kappa_{2}=KR_{2}-\pi j/2$, and we have defined 
\begin{widetext}
\begin{multline}\label{djdef}
\mathfrak{D}_j^\pm= \mbox{Im}\left[H_{j-1/2}^{(1)}(\rho_1)H_{j\mp 1/2}^{(2)}(\rho_2)\pm H_{j+1/2}^{(1)}(\rho_1)H_{j\pm 1/2}^{(2)}(\rho_2)\right] \\
=-J_{j-1/2}(\rho_1)Y_{j\mp 1/2}(\rho_2)+Y_{j-1/2}(\rho_1)J_{j\mp 1/2}(\rho_2)\mp J_{j+1/2}(\rho_1)Y_{j\pm 1/2}(\rho_2)\pm Y_{j+1/2}(\rho_1)J_{j\pm 1/2}(\rho_2),
\end{multline}
with $J_\nu(\rho)$ [$\,Y_\nu(\rho)\,$] the Bessel functions of the first [second] kind. The reflection and transmission amplitudes are
\begin{multline}
  r_{j}(E)=e^{-2i\kappa_{2}}\left(\eta\mathfrak{D}_{j}^-+i\mathfrak{D}_{j}^+\right)^{-1}\\ \times\left\{H_{j-1/2}^{(2)}(\rho_2)\left[H_{j-1/2}^{(1)}(\rho_1)+i{\eta}H_{j+1/2}^{(1)}(\rho_1)\right]-H_{j-1/2}^{(1)}(\rho_2)\left[H_{j-1/2}^{(2)}(\rho_1)+i{\eta}H_{j+1/2}^{(2)}(\rho_1)\right]-\eta\mathfrak{D}_{j}^--i\mathfrak{D}_{j}^+\right\},
\end{multline}
\end{widetext}
and
\begin{equation}\label{tjeta}
  t_{j}(E)=\frac{ 4{\eta}e^{iK(R_1-R_2)} }{ {\pi}k\sqrt{R_1R_2}\left(\eta\mathfrak{D}_{j}^-+i\mathfrak{D}_{j}^+\right) }.
\end{equation}
Defining $T_j\equiv |t_j(E)|^2$, we obtain Eq.\ (\ref{tjmodma}) of the main text. Notice that $T_j$ depends solely on $\mu_0=E-U_0$, as $t_j(E)$ is affected by $\mu_\infty=E-U_\infty$ only via a phase factor. It is also insensitive to the doping sign $\eta=\pm 1$, what corresponds to the particle-hole symmetry.

For the undoped-disk limit ($k\rightarrow 0$), Eq.\ (\ref{djdef}) leads to the asymptotic form
\begin{equation}
  \eta\mathfrak{D}_{j}^-+i\mathfrak{D}_{j}^+\approx \frac{2\eta}{{\pi}k\sqrt{R_1R_2}}\left[\left(\frac{R_1}{R_2}\right)^{j}+\left(\frac{R_2}{R_1}\right)^{j}\right].
\end{equation}
Substituting the above into Eq.\ (\ref{tjeta}) we obtain $T_j=|t_j(E\!\rightarrow\!U_0)|^2$ as given by Eq.\ (\ref{tjlambdadsk}) of the main text. Hence, the correspondence between the mode matching for angular-momentum eigenstates and the conformal mapping technique for the disk in undoped graphene is established.


\begin{thebibliography}{99}
\bibitem{TraInv}
The translational invariance along the $y$-axis is satisfied for an
infinitely-wide strip or for periodic boundary conditions of 
Ref.\ \cite{Kat06a}. In case of a confined geometry that does not mix the 
valleys, one can use $\Psi(x,y)=\chi_\theta(x)e^{ik_yy}+\chi_{-\theta}(x)e^{-ik_yy}$ 
for the mode-matching, and find that Eq.\ (\ref{tthphi}) remains unchanged.
\bibitem{Dircon}
Zero-energy solution of the Dirac equation in two dimensions has the general 
form $\Psi=[\Psi_A(x+iy),\Psi_B(x-iy)]^{T}$, where the spinor components 
$\Psi_A$ and $\Psi_B$ are holomorphic functions. 
For the strip geometry, infinite-mass boundary conditions
($\Psi_A|_{y=0}=\Psi_B|_{y=0}$, $\Psi_A|_{y=\mathcal{W}}=-\Psi_B|_{y=\mathcal{W}}$) are 
satisfied after applying the transformation $\Psi_A=\Psi_A(z(w))$, 
$\Psi_B=\Psi_B(z(w^\star))$, with $z(w)\equiv x(u+iv)+iy(u+iv)$ an analytical 
function that turns the coordinate system $(u,v)$ into $(x,y)$. The case of
antiperiodic b.c.\ is discussed in Section \ref{corbdi}.
\bibitem{perifoo}
In the special case of antiperiodic b.c., $s=8$ for all the modes. For
periodic b.c.\ considered in Ref.\ \cite{Kat06a} the lowest, fully transmitted
mode has only a fourfold (spin and valley) degeneracy.
\bibitem{match2deg}
We took $\Phi_l^\mathrm{I}(r)=H_l^{(2)}(Kr)+r_lH_l^{(1)}(Kr)$ for $r>R_2$,
$\Phi_l^\mathrm{II}(r)=a\mathcal{C}_l^{(1)}(kr)+b\mathcal{C}_l^{(2)}(kr)$ with 
$\mathcal{C}_l^{(1,2)}$ given by Eq.\ (\ref{calcdef}) for $R_2>r>R_1$,
$\Phi_l^\mathrm{III}(r)=t_lH_l^{(2)}(Kr)$ for $R_1>r$, and solved the matching
conditions 
$$
\Phi_l^\mathrm{I}(R_2)=\Phi_l^\mathrm{II}(R_2), \ \ \ \ \ \ 
\Phi_l^\mathrm{II}(R_1)=\Phi_l^\mathrm{III}(R_1),
$$
$$
\left.\frac{d\Phi_l^\mathrm{I}}{dr}\right|_{R_2}=
\left.\frac{d\Phi_l^\mathrm{II}}{dr}\right|_{R_2}, \ \ \ \ 
\left.\frac{d\Phi_l^\mathrm{II}}{dr}\right|_{R_1}=
\left.\frac{d\Phi_l^\mathrm{III}}{dr}\right|_{R_1}.
$$
\bibitem{land2deg}
We use the Landauer formula $G=(2e^2/h)\sum_{l=-\infty}^{\infty}T_l$, and 
$F=\sum_{l=-\infty}^{\infty}T_l(1-T_l)/\sum_{l=-\infty}^{\infty}T_l$. Note the twofold
degeneracy $T_{l}=T_{-l}$ for $l\neq 0$.
\bibitem{KirGscl}
The semiclassical conductance for nonrelativistic electrons in the Corbino disk is $G_{\mbox{\scriptsize s-cl}}=(4e^2/h)(l_1+1/2)$ with $l_1=\mbox{int}\left[(k^2R_1^2+1/4)^{1/2}\right]$, see Ref.\ \cite{Kir94}. It leads to $G\propto kR_1$ for $kR_1\gg 1$.
\bibitem{infdop2deg}
For a~2DEG and the limit of $K\gg k$ we find from Eq.\ (\ref{ttl2deg}) that $T_l\approx 64(\pi^2K^2R_1R_2)^{-1}|\mathcal{C}_l^{(1)}(kR_2)\mathcal{C}_l^{(2)}(kR_1)-\mathcal{C}_l^{(2)}(kR_2)\mathcal{C}_l^{(1)}(kR_1)|^{-2}$, which vanishes as $(1/K)^2$.
\bibitem{simcocdot}
For the half-Corbino disk ($\vartheta=\pi$) we have fixed the width of the 
wide lead attached from the right (see Fig.\ \ref{set2term}a) at 
$W_\infty=560a$ (corresponding to $N_\infty=213$ propagating modes for 
$\mu_\infty=\tau/2$). The inner radius is varied in the range 
$R_1=30\div 150a$ ($N_1=23\div 114$ modes). The remaining parameters are the 
outer radius $R_2=160a$ (the diameter $2R_2$ is noticeably smaller than 
$W_\infty$ to model the bulk-lead regime), 
and the edge-atoms chemical potential $\mu_A=-\mu_B=0.01\tau$.
For the quantum dot with circular edges (see Fig.\ \ref{set2term}b), 
we have fixed the lead radius at $R_1=40a$ ($N_1=30$ for 
$\mu_\infty=\tau/2$), the edge radius is varied in the range 
$24\sqrt{3}a<R_2<85\sqrt{3}a$, and the staggered-potential amplitude outside
the dot is varied in the range $|\mu_{A,B}|=0.1\div 0.8\tau$. We find that the
results are insensitive to the confinement strength.
\bibitem{simribbon}
The chemical potential $\mu_j=0$ in the undoped region (white areas in Fig.\ 
\ref{set4term}), except for the outermost edge atoms, where we put 
$\mu_j=\mu_{A,B}$, with $\mu_A=-\mu_B=0.01\div 0.1\tau$. 
In the leads (shadow areas) $\mu_j=\mu_\infty=0.4\tau$,
what corresponds to $60$ propagating modes (for the lead width fixed at 
$W=200a$). The sample area length $L$ is
varied, but kept equal to the width of undoped leads.
\bibitem{pdifsqu}
We suppose the current density $\mathbf{j}=\sigma_0\mathbf{E}=-\sigma_0\nabla
\varphi$, where $\varphi(x,y)$ is an electrostatic potential. The conductance 
$G=I/(\varphi_{x=L/2}-\varphi_{x=-L/2})$ is found by analytical integration of the 
current passing the vertical symmetry axis of the system ($x=0$).
\bibitem{simcisq}
Parameters of the nanoribbon-like system studied numerically are $L=96\sqrt{3}
a$, $W_\infty=560a$, $\mu_\infty=\tau/2$, $|\mu_{A,B}|=0.01\tau$. The narrow-lead 
width is varied in the range $W/a=40\div 300$.
\end{thebibliography}
\end{document}